\documentstyle[epsf]{mn}

\newcommand{\sA}{{\cal A}}
\newcommand{\sB}{{\cal B}}
\newcommand{\sC}{{\cal C}}
\newcommand{\sD}{{\cal D}}
\newcommand{\sI}{{\cal I}}
\newcommand{\sL}{{\cal L}}
\newcommand{\sR}{{\cal R}}
\newcommand{\sT}{{\cal T}}

\newcommand{\be}{\mbox{\boldmath$e$}}
\newcommand{\bff}{\mbox{\boldmath$f$}}
\newcommand{\br}{\mbox{\boldmath$r$}}
\newcommand{\bu}{\mbox{\boldmath$u$}}
\newcommand{\bv}{\mbox{\boldmath$v$}}
\newcommand{\bA}{\mbox{\boldmath$A$}}
\newcommand{\bB}{\mbox{\boldmath$B$}}
\newcommand{\bF}{\mbox{\boldmath$F$}}
\newcommand{\bqdot}{{\mbox{\boldmath$\dot q$}}}

\title[Non-linear fluid dynamics of eccentric discs]
{Non-linear fluid dynamics of eccentric discs}

\author[G. I. Ogilvie]
  {G. I. Ogilvie$^{1,2}$\\
  $^1$Institute of Astronomy, University of Cambridge, Madingley Road,
  Cambridge CB3 0HA\\
  $^2$Department of Applied Mathematics and Theoretical Physics,
  University of Cambridge, Silver Street, Cambridge CB3 9EW}

\begin{document}

\maketitle

\label{firstpage}

\begin{abstract}
  A new theory of eccentric accretion discs is presented.  Starting
  from the basic fluid-dynamical equations in three dimensions, I
  derive the fundamental set of one-dimensional equations that
  describe how the mass, angular momentum and eccentricity vector of a
  thin disc evolve as a result of internal stresses and external
  forcing.  The analysis is asymptotically exact in the limit of a
  thin disc, and allows for slowly varying eccentricities of arbitrary
  magnitude.  The theory is worked out in detail for a Maxwellian
  viscoelastic model of the turbulent stress in an accretion disc.
  This generalizes the conventional alpha viscosity model to account
  for the non-zero relaxation time of the turbulence, and is
  physically motivated by a consideration of the nature of
  magnetohydrodynamic turbulence.  It is confirmed that circular discs
  are typically viscously unstable to eccentric perturbations, as
  found by Lyubarskij, Postnov \& Prokhorov, if the conventional alpha
  viscosity model is adopted.  However, the instability can usually be
  suppressed by introducing a sufficient relaxation time and/or bulk
  viscosity.  It is then shown that an initially uniformly eccentric
  disc does not retain its eccentricity as had been suggested by
  previous analyses.  The evolutionary equations should be useful in
  many applications, including understanding the origin of planetary
  eccentricities and testing theories of quasi-periodic oscillations
  in X-ray binaries.
\end{abstract}


\section{Introduction}

\subsection{Background}

In a thin accretion disc the stresses due to collective effects are
relatively weak and the motion of the gas is nearly ballistic.  Since
the gravitational field is dominated by the central mass, the
principal motion consists of Keplerian orbits.  In the classical
theory of accretion discs (e.g. Pringle 1981) these orbits are assumed
to be circular and coplanar.  Not only is this the simplest situation,
it is also the expected outcome of the enhanced dissipation of energy
that would result from initially misaligned or eccentric orbits.

Nevertheless, there are strong observational and theoretical grounds
for believing that accretion discs in various situations are not flat
but warped (e.g. Ogilvie 2000 and references therein).  Such a
situation can be caused by external forcing or can arise spontaneously
through instabilities of an initially flat disc.  Similarly, there are
equally good reasons for investigating the possibility of eccentric
discs.  (The general case of a warped {\it and\/} eccentric disc
deserves attention but is beyond the scope of the present paper.)

The recent discovery of extrasolar planets orbiting main-sequence
stars (Mayor \& Queloz 1995; Marcy, Cochran \& Mayor 2000) is a major
development in the history of astronomy.  The distribution of orbital
elements of the planets discovered to date poses some important
challenges to theoretical models of planetary formation and dynamics.
In particular, all of the planets with semi-major axes greater than
0.2 AU have significant eccentricities, several exceeding 0.5.  One of
the leading contenders for driving eccentricity is the tidal
interaction between the planet and the disc from which it forms (e.g.
Lubow \& Artymowicz 2000).

A companion object on an eccentric orbit has a complicated tidal
interaction with the disc.  This is of considerable importance in
connection with planetary rings and young binary systems, in addition
to extrasolar planets.  In the classical theory (Goldreich \& Tremaine
1980), tightly wrapped density waves are launched at an array of
resonances and propagate some way through the disc before dissipating.
The resulting torques lead to evolution of the orbit of the companion
and of the surface density of the disc.  In addition, the companion
may couple to global, low-frequency eccentric motions of the disc
which are not described by the classical theory.  The importance of
such motions has been recognized by Tremaine (1998), and the methods
required to understand them will be presented in this paper.

Further arguments support the notion of eccentric discs.  Precessing
eccentric discs are understood to exist in superhump binaries as the
result of a tidal instability (e.g. Lubow 1991a and references
therein).  The general-relativistic apsidal precession of an eccentric
inner disc is one of the leading explanations for kilohertz
quasi-periodic oscillations in X-ray binaries (Stella, Vietri \&
Morsink 1999; Psaltis \& Norman 2000; but see Markovi\'c \& Lamb 2000
for a critical assessment).  A transitory eccentric disc may be
produced after tidal disruption of a star close to a black hole at the
centre of a galaxy (Gurzadyan \& Ozernoy 1979).  Many-body systems
with related dynamics include planetary rings, several of which have
small but accurately measured eccentricities (e.g. Borderies,
Goldreich \& Tremaine 1983 and references therein), and the nucleus of
the galaxy M31, for which Tremaine (1995) has proposed the model of an
eccentric Keplerian disc.

The existing theoretical work on eccentric discs consists of
analytical studies and numerical simulations, based almost exclusively
on a two-dimensional description of the disc.  Numerous authors have
treated small eccentric perturbations of an initially circular disc as
a special case of wave modes.  Kato (1983) showed that global,
low-frequency eccentric modes exist in an inviscid disc.  A slow
precession of the modes occurs owing to pressure forces.  In a
strictly inviscid disc, eccentric modes exist that have non-trivial
vertical structure (Okazaki \& Kato 1985), although these are not
expected to be important when viscosity is included.  A similar, modal
description was used by Hirose \& Osaki (1993) to describe superhumps
in SU UMa binaries.  Ostriker, Shu \& Adams (1992) considered the
near-resonant excitation of eccentric density waves in a
self-gravitating disc due to a companion on an eccentric orbit.  Lee
\& Goodman (1999) used novel techniques to analyse non-linear
eccentric density waves in the tight-winding limit, for a
self-gravitating disc.

Much effort has been devoted to explaining the superhump phenomenon in
terms of a precessing, eccentric disc.  Lubow (1991a) showed that a
large-scale eccentric perturbation can grow through coupling with the
tidal potential in a circular binary system.  The coupling involves
waves that are launched at eccentric Lindblad resonances, which are
present only in tidally truncated discs in binaries of extreme mass
ratio, as in SU UMa systems.  Two-dimensional numerical simulations
have been performed by Whitehurst (1988), Hirose \& Osaki (1990),
Whitehurst \& King (1991), Lubow (1991b), Whitehurst (1994) and Murray
(1996, 1998, 2000).

More general studies are of direct relevance to the present paper.
Borderies et al. (1983) derived evolutionary equations for slightly
eccentric planetary rings subject to quadrupolar and tidal forces,
self-gravitation and viscosity.  Syer \& Clarke (1992, 1993) used the
Gauss perturbation equations of celestial mechanics to argue that,
contrary to the assumptions of classical accretion disc theory, an
isolated viscous disc will retain its initial eccentricity
indefinitely.  Their numerical simulations supported this idea.  Most
recently, Lyubarskij, Postnov \& Prokhorov (1994) derived evolutionary
equations for a viscous accretion disc in which the orbits are
ellipses sharing a common longitude of periastron.  They confirmed the
result of Syer \& Clarke (1993) but only in the sense of a singular
solution of the time-dependent equations, finding that an initially
circular disc is viscously unstable to eccentric perturbations if the
usual alpha viscosity model is adopted.

The aims of the present paper are to unify these previous treatments
and to extend them in a number of important ways.  A new set of
evolutionary equations for an eccentric disc will be derived.  The
analysis of Lyubarskij et al.  (1994) will be extended to allow for
precession of the orbits, which can never be avoided and is the
dominant feature in the pressure-driven modes of Kato (1983).  The
earlier modal description will be extended to allow for arbitrary
eccentricities, and the effects of turbulent stresses and radiation.
The analysis will also include important three-dimensional effects,
which have been neglected in previous studies.  In addition, the
conventional alpha viscosity model will be generalized into a
Maxwellian viscoelastic model, which is a physically motivated and
more realistic description of the turbulent stress in an accretion
disc.

\subsection{Continuum celestial mechanics}

As an introductory exercise, consider the problem of a test body
orbiting in the gravitational field of a central object of mass $M$,
but subject to a perturbing force per unit mass $\bff$.  Its equation
of motion is
\begin{equation}
  {{{\rm d}\bu}\over{{\rm d}t}}=-{{GM}\over{r^2}}\,\be_r+\bff,
  \label{gauss_motion}
\end{equation}
where $\bu={\rm d}\br/{\rm d}t$ is the velocity, with $\br=r\,\be_r$
being the position vector with respect to the central object.  Since
inclined orbits are beyond the scope of this paper, assume that the
orbit and the perturbing force lie in the $xy$-plane.

The traditional method of determining the rate of change of the
osculating orbital elements of the body involves evaluating the
disturbing function and applying the Gauss perturbation equations of
celestial mechanics (e.g. Brouwer \& Clemence 1961).  A more compact
derivation uses the {\it eccentricity vector\/} $\be$, which lies in the
plane of the orbit and may be defined by (Eggleton, Kiseleva \& Hut
1998; Lynden-Bell 2000)
\begin{equation}
  \bu={{GM}\over{h}}\,\be_z\times(\be_r+\be),
\end{equation}
or, equivalently,
\begin{equation}
  \be=-{{h}\over{GM}}\,\be_z\times\bu-\be_r,
\end{equation}
where
\begin{equation}
  h\,\be_z=\br\times\bu
\end{equation}
is the specific angular momentum.  The position and velocity are then
instantaneously equal to those of an elliptical Keplerian orbit of
semi-latus rectum
\begin{equation}
  \lambda={{h^2}\over{GM}}=r+\br\cdot\be
\end{equation}
and eccentricity $e=|\be|$, with $\be$ pointing towards periastron.

Substitution into the equation of motion (\ref{gauss_motion}) gives
the relations
\begin{equation}
  {{{\rm d}h}\over{{\rm d}t}}\,\be_z=\br\times\bff,
  \label{gauss_h}
\end{equation}
\begin{equation}
  h\,{{{\rm d}\be}\over{{\rm d}t}}=
  {{{\rm d}h}\over{{\rm d}t}}\,(\be_r+\be)-\lambda\,\be_z\times\bff,
  \label{gauss_e}
\end{equation}
which determine the evolution of the orbital elements and are
equivalent to the Gauss perturbation equations.

The problem to be addressed in this paper is how to derive a related
set of equations for a fluid disc in which $\be=\be(\lambda,t)$ varies
not only with time but also from one orbit to the next.  The
evolutionary equations will be partial differential equations (PDEs)
depending on one spatial variable, analogous to those obtained for
warped discs (Pringle 1992; Ogilvie 1999, 2000), and equally suitable
for semi-analytical or numerical implementation.

The perturbing force in a fluid disc includes a contribution from the
internal stress in addition to any external forcing.  The
instantaneous elliptical Keplerian motion associated with the function
$\be(\lambda,t)$ determines (part of) the velocity gradient tensor in
the disc.  An appropriate theory, satisfying the basic principles of
continuum mechanics, is then required to relate the stress tensor to
the velocity gradient tensor.

\subsection{Modelling the turbulent stress}

The simplest assumption, made in almost all theoretical work on
accretion discs, is that the turbulent stress may be treated as an
isotropic effective viscosity in the sense of the Navier-Stokes
equation.\footnote{Note the crucial distinction between isotropy of
  the stress and isotropy of the effective viscosity.} The stress
tensor may then be written as
\begin{equation}
  {\bf T}=\mu\left[\nabla\bu+(\nabla\bu)^{\rm T}\right]+
  (\mu_{\rm b}-{\textstyle{{2}\over{3}}}\mu)
  (\nabla\!\cdot\!\bu){\bf1},
\end{equation}
where $\mu$ is the effective (dynamic) shear viscosity and $\mu_{\rm
  b}$ the effective bulk viscosity (which is often omitted).  For
the usual case of a steady, flat and circular Keplerian disc, this
assumption suffices to generate the stress component $T_{R\phi}$ that
is required for accretion to proceed.  Together with the conventional
parametrization of the viscosity coefficient, this constitutes the
alpha viscosity model (Shakura \& Sunyaev 1973), which was very
successful in connecting theory and observations during a period when
the theory was incomplete.

In recent years it has become widely accepted that the turbulent
stress in most, if not all, accretion discs is magnetohydrodynamic
(MHD) in origin, resulting from the non-linear development of the
magnetorotational instability (Balbus \& Hawley 1998).  The turbulent
stress tensor is
\begin{equation}
  {\bf T}=\bigg\langle{{\bB'\bB'}\over{4\pi}}-\rho\bu'\bu'-
  \rho'(\bu\bu'+\bu'\bu+\bu'\bu')\bigg\rangle,
  \label{turbulent}
\end{equation}
where the prime denotes a turbulent fluctuation from the mean value of
a quantity, and the magnetic pressure fluctuation has been considered
to be combined with the gas pressure fluctuation.  Numerical
simulations of the turbulence offer the possibility of measuring the
stress and testing the validity of the alpha viscosity model in more
general circumstances.  In fact very few studies have been directed
towards this aim (Abramowicz, Brandenburg \& Lasota 1996; Torkelsson
et al. 2000).

The most obvious deficiency of the alpha viscosity model is that it
requires the stress to respond instantaneously to a change in the rate
of strain.  In an eccentric disc the strain field experienced by a
fluid element changes during the course of each orbit.  There is a
strong suspicion that the assumption of an instantaneous response may
be partially responsible for the eccentric instability described by
Lyubarskij et al. (1994).  In reality the turbulence is expected to
have a non-zero relaxation time comparable (on dimensional grounds) to
the orbital period.

It is therefore proposed that the stress tensor satisfies the equation
\begin{equation}
  {\bf T}+\tau{\sD}{\bf T}=\mu\left[\nabla\bu+(\nabla\bu)^{\rm T}\right]+
  (\mu_{\rm b}-{\textstyle{{2}\over{3}}}\mu)
  (\nabla\!\cdot\!\bu){\bf1},
  \label{stress}
\end{equation}
where $\tau$ is the relaxation time and $\sD$ a linear operator defined by
\begin{equation}
  \sD{\bf T}=\partial_t{\bf T}+\bu\!\cdot\!\nabla{\bf T}-
  (\nabla\bu)^{\rm T}\cdot{\bf T}-{\bf T}\!\cdot\!\nabla\bu+
  2(\nabla\!\cdot\!\bu){\bf T}.
\end{equation}
This model (although with $\nabla\!\cdot\!\bu=0$ for an incompressible
fluid) is used in rheology as a simple model of a viscoelastic medium,
known as the {\it upper-convected Maxwell fluid}.  The relaxation time
was introduced by Clerk Maxwell (1866) in his study of the viscosity
of gases.  The Maxwellian viscoelastic fluid allows for a continuum of
behaviour from an elastic solid (on time-scales short compared to
$\tau$) to a viscous fluid (on time-scales long compared to $\tau$).
The operator $\sD$ is one of a family of convective derivative
operators, acting on second-rank tensors, that satisfy material frame
indifference, which is a fundamental principle of continuum mechanics.
In the context of rheology, equation (\ref{stress}) can be justified
either by abstract principles of continuum mechanics, or from a
kinetic theory of polymer molecules that satisfy Hooke's law (Bird,
Armstrong \& Hassager 1987; Bird, Curtiss \& Armstrong 1987).

There are good reasons to suppose that a viscoelastic model may offer
a reasonable description of MHD turbulence in accretion discs.
Magnetic field lines have tension and support Alfv\'en waves similar
to waves on an elastic string.  There is a close analogy between the
tangled magnetic field lines in a turbulent medium and the tangled
polymer molecules in a viscoelastic fluid.  Indeed, the equation
satisfied by the dyadic tensor $\bB\bB$ in ideal MHD is
precisely\footnote{I am indebted to Prof. M. R. E. Proctor for pointing
  this out.}
\begin{equation}
  \sD(\bB\bB)=0,
\end{equation} 
and $\langle\bB'\bB'/4\pi\rangle$ is arguably the most important
contribution to the total turbulent stress tensor (\ref{turbulent}).

This accounts for the `elastic' aspect of MHD turbulence.  The
`viscous' part of equation (\ref{stress}) can then be related
informally to the (less well understood) dissipative and non-linear
aspects of the turbulence.  Viscoelastic models have been proposed for
hydrodynamic turbulence (Crow 1968) but the case for MHD turbulence is
arguably stronger.

This hypothesis is also related to the `causal viscosity' model used
in the context of accretion disc boundary layers to ensure that
information is not propagated faster than the sound speed (Kley \&
Papaloizou 1997).  In that context, only a single component of ${\bf
  T}$ is treated and the full expression for $\sD$ is not used.  Here,
however, it is essential to treat the stress as a tensor and to ensure
that the equations satisfy material frame indifference.

Note that one easily recovers the conventional alpha viscosity model
by taking the limit $\tau\to0$.

It is of interest to work out the implications of this model for a
steady, flat and circular Keplerian disc.  The horizontal stress
components are predicted to be
\begin{equation}
  T_{RR}=0,\qquad
  T_{R\phi}=-{{3}\over{2}}\mu\Omega,\qquad
  T_{\phi\phi}={{9}\over{2}}{\rm We}\,\mu\Omega,
\end{equation}
where $\Omega$ is the angular velocity and
\begin{equation}
  {\rm We}=\Omega\tau
\end{equation}
the {\it Weissenberg number}, which is expected to be of order unity.
A comparison with the results of MHD simulations by Hawley, Gammie \&
Balbus (1995) suggests that the Maxwellian viscoelastic model performs
rather well in predicting the turbulent stress tensor
(\ref{turbulent}).  The results of Table~2 in that paper imply that
\begin{eqnarray}
  \lefteqn{(T_{RR},T_{R\phi},T_{Rz},T_{\phi\phi},T_{\phi z},T_{zz})}
  &\nonumber\\
  &&\propto(0.105,-1,0.018,1.520,-0.050,0.076),
\end{eqnarray}
whereas the viscoelastic model predicts
\begin{equation}
  (T_{RR},T_{R\phi},T_{Rz},T_{\phi\phi},T_{\phi z},T_{zz})\propto
  (0,-1,0,3{\rm We},0,0).
\end{equation}
Given that the numerical quantities are averaged over a limited time
interval and involve some statistical error, the agreement is good and
the results are consistent with ${\rm We}\approx0.5$.  It would be
valuable to conduct further simulations designed to measure the
relaxation time in a more direct way.

\subsection{Plan of the paper}

The remainder of this paper is organized as follows.  In Section~2 I
present a highly simplified analysis of an eccentric disc, with the
purpose of explaining the principles of the method without the
technical detail of the full model.  In Section~3 I set out the basic
equations and coordinate systems used in the rest of the paper.
Section~4 contains the bulk of the analysis.  I present an asymptotic
development of the equations appropriate for a thin disc, reducing the
problem to simpler units that are solved in turn to extract the
required one-dimensional evolutionary equations.  In Section~5 a
linearized theory for small eccentricity is described and compared
with the oversimple model of Section~2.  An illustrative,
time-dependent solution of the fully non-linear equations is given in
Section~6.  Finally, in Section~7, the results are summarized and a
comparison with existing work is made.  Recommendations for
applications of this work are also given.

\section{An oversimple model}

\subsection{Analysis}

In the following sections a detailed model of a thin, eccentric
accretion disc will be developed, consisting of a fully non-linear
asymptotic analysis of the fluid-dynamical equations in three
dimensions, including dissipation of energy and radiative transport.
Before setting out this theory, I present a grossly simplified model
(two-dimensional, linear and barotropic) that is not recommended for
further use but nevertheless illustrates some of the important
principles without many of the technical details of the full
calculation.

Suppose that the fluid is strictly two-dimensional and obeys the
equation of mass conservation,
\begin{equation}
  (\partial_t+\bu\!\cdot\!\nabla)\rho=-\rho\nabla\!\cdot\!\bu,
\end{equation}
and the equation of motion,
\begin{equation}
  \rho(\partial_t+\bu\!\cdot\!\nabla)\bu=-\rho\nabla\Phi-\nabla p+
  \nabla\!\cdot\!{\bf T},
\end{equation}
where $\rho$ is the density, $\bu$ the velocity, $p$ the pressure and
${\bf T}$ the turbulent stress tensor, which will be assumed to
satisfy equation (\ref{stress}).  The gravitational potential of the
central mass $M$ is $\Phi=-GM/R$, referred to plane polar coordinates
$(R,\phi)$.  The self-gravitation of the fluid is neglected.

Much use has been made of the analogy between the dynamics of a
two-dimensional fluid and that of a thin disc.  However, the analogy
is not a formal one except under special circumstances (and never for
an eccentric disc, as will be seen).  In fact, a two-dimensional model
corresponds to an infinite cylinder rather than a thin disc, although
the gravitational potential cannot be interpreted in this sense.
However, $\rho$, $p$ and $\mu$ in the two-dimensional theory might be
regarded as approximately equivalent to the corresponding vertically
integrated quantities in the three-dimensional theory.  To emphasize
this, in this section the density will be written as $\Sigma$ (surface
density) instead of $\rho$.  To avoid a consideration of the energy
equation at this stage, suppose that the fluid is barotropic, so that
$p=p(\Sigma)$, $\mu=\mu(\Sigma)$ and $\mu_{\rm b}=\mu_{\rm b}(\Sigma)$
are given functions.

Let $(u,v)$ be the polar components of $\bu$.  Then one has
\begin{equation}
  \left(\partial_t+u\partial_R+{{v}\over{R}}\partial_\phi\right)\Sigma=
  -{{\Sigma}\over{R}}\left[\partial_R(Ru)+\partial_\phi v\right],
  \label{simple_sigma}
\end{equation}
\begin{eqnarray}
  \lefteqn{\Sigma\left[
  \left(\partial_t+u\partial_R+{{v}\over{R}}\partial_\phi\right)u-
  {{v^2}\over{R}}\right]=-\Sigma{{GM}\over{R^2}}-\partial_Rp}&\nonumber\\
  &&+{{1}\over{R}}\partial_R(RT_{RR})+{{1}\over{R}}\partial_\phi T_{R\phi}-
  {{T_{\phi\phi}}\over{R}},
  \label{simple_u}
\end{eqnarray}
\begin{eqnarray}
  \lefteqn{\Sigma\left[
  \left(\partial_t+u\partial_R+{{v}\over{R}}\partial_\phi\right)v+
  {{uv}\over{R}}\right]=-{{1}\over{R}}\partial_\phi p}&\nonumber\\
  &&+{{1}\over{R^2}}\partial_R(R^2T_{R\phi})+
  {{1}\over{R}}\partial_\phi T_{\phi\phi},
  \label{simple_v}
\end{eqnarray}
with
\begin{eqnarray}
  \lefteqn{T_{RR}+\tau\left\{
  \left(\partial_t+u\partial_R+{{v}\over{R}}\partial_\phi\right)T_{RR}+
  {{2}\over{R}}(u+\partial_\phi v)T_{RR}\right.}&\nonumber\\
  &&\left.-{{2}\over{R}}(\partial_\phi u)T_{R\phi}\right\}=
  2\mu\partial_Ru\nonumber\\
  &&+(\mu_{\rm b}-{\textstyle{{2}\over{3}}}\mu)
  {{1}\over{R}}\left[\partial_R(Ru)+\partial_\phi v\right],
\end{eqnarray}
\begin{eqnarray}
  \lefteqn{T_{R\phi}+\tau\left\{
  \left(\partial_t+u\partial_R+{{v}\over{R}}\partial_\phi\right)T_{R\phi}-
  R\partial_R\left({{v}\over{R}}\right)T_{RR}\right.}&\nonumber\\
  &&\left.
  +{{1}\over{R}}\left[\partial_R(Ru)+\partial_\phi v\right]T_{R\phi}-
  {{1}\over{R}}(\partial_\phi u)T_{\phi\phi}\right\}\nonumber\\
  &&=\mu\left[R\partial_R\left({{v}\over{R}}\right)+
  {{1}\over{R}}\partial_\phi u\right],
\end{eqnarray}
\begin{eqnarray}
  \lefteqn{T_{\phi\phi}+\tau\left\{
  \left(\partial_t+u\partial_R+{{v}\over{R}}\partial_\phi\right)T_{\phi\phi}-
  2R\partial_R\left({{v}\over{R}}\right)T_{R\phi}\right.}&\nonumber\\
  &&\left.+2(\partial_Ru)T_{\phi\phi}\right\}=
  2\mu\left({{u}\over{R}}+{{1}\over{R}}\partial_\phi v\right)\nonumber\\
  &&+(\mu_{\rm b}-{\textstyle{{2}\over{3}}}\mu)
  {{1}\over{R}}\left[\partial_R(Ru)+\partial_\phi v\right].
\end{eqnarray}

Let the basic state be an axisymmetric, circular disc that evolves on
the slow, viscous time-scale.  Adopt a system of units such that the
radius of the disc and the characteristic orbital time-scale are
$O(1)$.  The slow evolution is captured by a slow time coordinate
$T=\epsilon^2t$, where the small parameter $\epsilon$ is a
characteristic value of the angular semi-thickness $H/R$ of the disc.
For the basic state, introduce the expansions
\begin{eqnarray}
  u&=&\epsilon^2u_2(R,T)+O(\epsilon^4),\\
  v&=&R\Omega(R)+\epsilon^2v_2(R,T)+O(\epsilon^4),\\
  \Sigma&=&\Sigma_0(R,T)+O(\epsilon^2),\\
  p&=&\epsilon^2\left[p_0(R,T)+O(\epsilon^2)\right],\\
  \mu&=&\epsilon^2\left[\mu_0(R,T)+O(\epsilon^2)\right],\\
  \mu_{\rm b}&=&\epsilon^2\left[\mu_{{\rm b}0}(R,T)+O(\epsilon^2)\right],\\
  {\bf T}&=&\epsilon^2\left[{\bf T}_0(R,T)+O(\epsilon^2)\right],
\end{eqnarray}
where
\begin{equation}
  \Omega=\left({{GM}\over{R^3}}\right)^{1/2}
\end{equation}
is the Keplerian angular velocity.
Substitution into equations (\ref{simple_sigma})--(\ref{simple_v}) leads to
\begin{equation}
  (\partial_T+u_2\partial_R)\Sigma_0=-{{\Sigma_0}\over{R}}\partial_R(Ru_2),
\end{equation}
\begin{equation}
  -2\Sigma_0\Omega v_2=-\partial_Rp_0-{{T_{\phi\phi0}}\over{R}},
\end{equation}
\begin{equation}
  {{1}\over{2}}\Sigma_0\Omega u_2=
  {{1}\over{R^2}}\partial_R(R^2T_{R\phi0}),
\end{equation}
with
\begin{eqnarray}
  T_{RR0}&=&0,\\
  T_{R\phi0}&=&-{{3}\over{2}}\mu_0\Omega,\\
  T_{\phi\phi0}&=&{{9}\over{2}}{\rm We}\,\mu_0\Omega,
\end{eqnarray}
and this leads to the well known evolutionary equation for the surface
density (e.g. Pringle 1981),
\begin{equation}
  \partial_T\Sigma_0={{3}\over{R}}\partial_R
  \left[R^{1/2}\partial_R\left(R^{1/2}\mu_0\right)\right].
  \label{simple_mass}
\end{equation}

Now consider slowly varying, one-armed linear perturbations of the basic
state, such that the Eulerian perturbation of $u$, say, is
\begin{equation}
  {\rm Re}\left[u'(R,T)\,{\rm e}^{-{\rm i}\phi}\right].
\end{equation}
For the perturbations, introduce the expansions
\begin{eqnarray}
  u'&=&u_0'(R,T)+\epsilon^2u_2'(R,T)+O(\epsilon^4),\\
  v'&=&v_0'(R,T)+\epsilon^2v_2'(R,T)+O(\epsilon^4),\\
  \Sigma'&=&\Sigma_0'(R,T)+O(\epsilon^2),\\
  p'&=&\epsilon^2\left[p_0'(R,T)+O(\epsilon^2)\right],\\
  \mu'&=&\epsilon^2\left[\mu_0'(R,T)+O(\epsilon^2)\right],\\
  \mu_{\rm b}'&=&\epsilon^2\left[\mu_{{\rm b}0}'(R,T)+
  O(\epsilon^2)\right],\\
  {\bf T}'&=&\epsilon^2\left[{\bf T}_0'(R,T)+O(\epsilon^2)\right].
\end{eqnarray}
In this linear analysis, the overall amplitude of the perturbations is
of course arbitrary, and the above scaling is chosen for convenience.

Equations (\ref{simple_u}) and (\ref{simple_v}) at $O(1)$ yield
\begin{equation}
  \Sigma_0(-{\rm i}\Omega u_0'-2\Omega v_0')=0,
\end{equation}
\begin{equation}
  \Sigma_0(-{\rm i}\Omega v_0'+{{1}\over{2}}\Omega u_0')=0,
\end{equation}
with the general solution
\begin{eqnarray}
  u_0'&=&{\rm i}R\Omega E,\\
  v_0'&=&{{1}\over{2}}R\Omega E,
\end{eqnarray}
where $E(R,T)$ is a complex function to be determined.  It is
easily verified that this solution corresponds to a small eccentric
perturbation, with complex eccentricity $E=e\,{\rm e}^{{\rm i}\omega}$
corresponding to eccentricity $e(R,T)$ and longitude of periastron
$\omega(R,T)$.  Equivalently, $E=e_x+{\rm i}e_y$, where $(e_x,e_y)$
are the Cartesian components of the eccentricity vector.

Equation (\ref{simple_sigma}) at $O(1)$ then yields the density perturbation,
\begin{equation}
  \Sigma_0'=R\partial_R(\Sigma_0E).
\end{equation}
The pressure and viscosity perturbations follow from the barotropic
relations, e.g.
\begin{equation}
  p_0'=\left({{{\rm d}p}\over{{\rm d}\Sigma}}\right)\Sigma_0'.
\end{equation}

Equations (\ref{simple_u}) and (\ref{simple_v}) at $O(\epsilon^2)$
contain $u_2'$ and $v_2'$, but these may be eliminated by taking an
appropriate linear combination.  This results in an evolutionary
equation for $E$, which may be written in the form
\begin{eqnarray}
  \lefteqn{2\Sigma_0R\Omega\partial_T E=
  -2\Sigma_0u_2\partial_R(R\Omega E)-
  \Omega\partial_R(R\Sigma_0u_2E)}&\nonumber\\
  &&-2{\rm i}R^{1/2}\Omega\partial_R(R^{1/2}\Sigma_0v_2E)+
  {{\rm i}\over{R^2}}\partial_R(R^2p_0')\nonumber\\
  &&-{{{\rm i}}\over{R}}\partial_R(RT_{RR0}')+
  2R^{-3/2}\partial_R(R^{3/2}T_{R\phi0}')-
  {{{\rm i}T_{\phi\phi0}'}\over{R}}.
  \label{simple_ecc}
\end{eqnarray}
The stress perturbations satisfy the equations
\begin{eqnarray}
  \lefteqn{T_{RR0}'+{\rm We}(-{\rm i}T_{RR0}'-2T_{R\phi0}E)}&\nonumber\\
  &&=2{\rm i}\mu_0\partial_R(R\Omega E)+
  {\rm i}(\mu_{{\rm b}0}-{\textstyle{{2}\over{3}}}\mu_0)R\Omega\partial_RE,
\end{eqnarray}
\begin{eqnarray}
  \lefteqn{T_{R\phi0}'+{\rm We}\left[
  {{3}\over{2}}T_{RR0}'-{\rm i}T_{R\phi0}'+
  {\rm i}R\partial_R(T_{R\phi0}E)-T_{\phi\phi0}E\right]}&\nonumber\\
  &&=-{{3}\over{2}}\left({{\partial\mu}\over{\partial\Sigma}}\right)
  R\Omega\partial_R(\Sigma_0E)+{{1}\over{2R}}\mu_0\partial_R(R^2\Omega E),
\end{eqnarray}
\begin{eqnarray}
  \lefteqn{T_{\phi\phi0}'+{\rm We}\left[
  3T_{R\phi0}'-{\rm i}T_{\phi\phi0}'-R^{5/2}\partial_R(R^{-3/2}E)T_{R\phi0}
  \right.}&\nonumber\\
  &&\left.+{\rm i}RE\partial_RT_{\phi\phi0}+
  2{\rm i}R^{3/2}\partial_R(R^{-1/2}E)T_{\phi\phi0}\right]\nonumber\\
  &&={\rm i}\mu_0\Omega E+
  {\rm i}(\mu_{{\rm b}0}-{\textstyle{{2}\over{3}}}\mu_0)R\Omega\partial_RE.
\end{eqnarray}

Thus one has obtained evolutionary equations for the surface density
(equation \ref{simple_mass}) and the complex eccentricity variable
(equation \ref{simple_ecc}).  In this linear theory the equations are
decoupled because the small eccentricity makes a negligible change to
the stresses that cause accretion.

\subsection{Evolution of eccentricity in a purely viscous fluid}

Equation (\ref{simple_ecc}) may be written in the form of a linear
evolutionary equation for the eccentricity,
\begin{equation}
  2\Sigma_0R^2\partial_T E=
  {\sA}R^2\partial_R^2E+{\sB}R\partial_RE+{\sC}E.
  \label{linear_e_eq}
\end{equation}
The general form of the complex coefficients $({\sA},{\sB},{\sC})$ is
complicated, but in the limit of a purely viscous disc, ${\rm
  We}\to0$, one obtains
\begin{equation}
  {\sA}=3\mu-3\Sigma\left({{{\rm d}\mu}\over{{\rm d}\Sigma}}\right)+
  (\mu_{\rm b}-{\textstyle{{2}\over{3}}}\mu)+
  {{{\rm i}\Sigma}\over{\Omega}}\left({{{\rm d}p}\over{{\rm d}\Sigma}}\right),
\end{equation}
\begin{eqnarray}
  \lefteqn{{\sB}={{12}\over{R\Omega}}\partial_R(R^2\Omega\mu)-
  3R\partial_R\mu-
  3\partial_R\left(R\Sigma{{{\rm d}\mu}\over{{\rm d}\Sigma}}\right)}
  &\nonumber\\
  &&+{{1}\over{R^2\Omega}}\partial_R
  \left[(\mu_{\rm b}-{\textstyle{{2}\over{3}}}\mu)R^3\Omega\right]+
  {{{\rm i}}\over{R^2\Omega}}\partial_R
  \left(R^3\Sigma{{{\rm d}p}\over{{\rm d}\Sigma}}\right),
\end{eqnarray}
\begin{equation}
  {\sC}=-2R\partial_R\mu+{{{\rm i}R}\over{\Omega}}\partial_Rp,
\end{equation}
where the subscript zeros have been omitted.

First, suppose that an initial condition is specified in which $E={\rm
  constant}$, so that the eccentricity is uniform and the ellipses are
aligned.  The initial rate of change of $E$ is determined by the
coefficient ${\sC}$.  In a steady disc far from the inner radius,
$\mu$ is nearly independent of $R$ and the real part of ${\sC}$ nearly
vanishes.  Therefore there is no initial viscous evolution of
eccentricity, as has been argued by Syer \& Clarke (1992, 1993) and
Lyubarskij et al. (1994).  However, the imaginary part of ${\sC}$ does
not vanish (except in the unlikely case that the pressure is
independent of $R$) and causes a differential precession of the
orbits.  Viscosity will then act on the twisted configuration.  It
follows that a steady, uniformly eccentric disc is {\it not\/} a
solution of the model.  This difference comes about because the
calculations of Syer \& Clarke and Lyubarskij et al. neglected the
differential precession caused by slightly non-Keplerian rotation
resulting from the radial pressure gradient.

Now consider a limit in which the eccentricity varies on a length-scale
$\ell\ll R$.  (One requires $\ell\gg H$, however, in order that the
ordering scheme adopted above remain valid.)  Then equation
(\ref{simple_ecc}) becomes approximately
\begin{eqnarray}
  \lefteqn{\partial_TE\approx{{1}\over{2\Sigma}}
  \left[3\mu-3\Sigma\left({{{\rm d}\mu}\over{{\rm d}\Sigma}}\right)+
  (\mu_{\rm b}-{\textstyle{{2}\over{3}}}\mu)\right.}&\nonumber\\
  &&\qquad\qquad\left.+{{{\rm i}\Sigma}\over{\Omega}}
  \left({{{\rm d}p}\over{{\rm d}\Sigma}}\right)\right]\partial_R^2E.
\end{eqnarray}
This is a linear diffusion equation with a complex diffusion
coefficient.  The imaginary part of the coefficient, which is due to
pressure, results in wave-like propagation of the eccentricity and to
the modes studied by Kato (1983).  The real part determines the
viscous diffusion of eccentricity.  If the real part of the diffusion
coefficient is negative, instability occurs.  {\it Within this
  oversimple model}, therefore, an initially circular disc is unstable
to developing eccentricity on small scales if
\begin{equation}
  {{{\rm d}\ln\mu}\over{{\rm d}\ln\Sigma}}>
  1+{{1}\over{3}}\left({{\mu_{\rm b}}\over{\mu}}-{{2}\over{3}}\right),
  \label{simple_criterion}
\end{equation}
i.e. if the vertically integrated viscosity increases sufficiently
rapidly with increasing surface density.  This criterion agrees with
the findings of Lyubarskij et al. (1994) if the bulk term (the term in
brackets) is neglected.  However, this analysis shows that a
sufficiently large bulk viscosity stabilizes the
disc.

Interestingly, equation (\ref{simple_criterion}) is identical to the
criterion for viscous overstability of short-wavelength {\it
  axisymmetric} modes in a two-dimensional disc (Kato 1978; see
Willerding 1998 for the effect of bulk viscosity).  This should not be
surprising because the local properties of waves in a thin disc are
almost independent of the azimuthal wavenumber when the radial
wavelength is much shorter than the azimuthal wavelength.  A related
result was also found by Goldreich \& Tremaine (1978), who examined
the damping effect of shear and bulk viscosity on density waves in
planetary rings.  They considered the case in which $\mu$ and
$\mu_{\rm b}$ were constant, but commented that viscosity might
produce growth rather than decay if $\mu$ and $\mu_{\rm b}$ were to
depend on $\Sigma$.  Note that the condition (\ref{simple_criterion})
is not related to the criterion for thermal-viscous instability, ${\rm
  d}\ln\mu/{\rm d}\ln\Sigma<0$.

In the alpha models (Shakura \& Sunyaev 1973) for gas-pressure
dominated discs, $\partial\ln\mu/\partial\ln\Sigma$ is equal to $5/3$
for a Thomson opacity law and $10/7$ for a Kramers opacity law.
Stability therefore requires $\mu_{\rm b}/\mu$ to exceed $8/3$ or
$41/21$, respectively.  Although bulk viscosity is not often discussed
in the context of alpha models, there is no reason to suppose that
accretion disc turbulence has a vanishing effective bulk viscosity.
In addition, it is known that radiation damping, which has been
neglected here, can mimic a bulk viscosity (Papaloizou \& Pringle
1977).  If the instability does occur, however, it will produce
short-wavelength variations that invalidate the theory developed here.
The non-linear outcome would have to be followed using numerical
simulations that allowed for the viscosity to be a function of the
surface density.  The non-linear development of the axisymmetric version
of the instability has been computed by, e.g., Papaloizou \& Stanley
(1986) and Kley, Papaloizou \& Lin (1993).

Starting from equation (\ref{simple_ecc}) it is possible to derive a
conservation law for eccentricity in the form\footnote{I am indebted
  to Dr S. H. Lubow for suggesting this calculation.}
\begin{eqnarray}
  \lefteqn{\partial_T(\Sigma R^2\Omega|E|^2)=
  {{1}\over{2R}}\partial_R\left[\Sigma{{{\rm d}p}\over{{\rm d}\Sigma}}R^3
  ({\rm i}E^*\partial_RE-{\rm i}E\partial_RE^*)\right]}&\nonumber\\
  &&+\hbox{viscous terms}.
\end{eqnarray}
This shows that, in the absence of viscosity, the eccentricity is
conserved in a certain sense.  The conserved quantity is related to
the `angular momentum deficit' which is conserved in the secular
theory of celestial mechanics (e.g. Laskar 1997).  The eccentricity
flux due to pressure effects is similar to the particle flux in
Schr\"odinger's equation.  The viscous terms are not necessarily
negative definite, indicating the possibility of instability as
described above.

\subsection{Effect of a non-zero relaxation time}

The general form of the coefficient ${\sA}$ for a Maxwellian
viscoelastic fluid is
\begin{eqnarray}
  \lefteqn{{\sA}=(1-{\rm i}{\rm We})^{-1}\left\{3\mu-
  3\Sigma\left({{{\rm d}\mu}\over{{\rm d}\Sigma}}\right)+
  (\mu_{\rm b}-{\textstyle{{2}\over{3}}}\mu)\right.}&\nonumber\\
  &&\left.-\left({{3{\rm i}{\rm We}}\over{1-{\rm i}{\rm We}}}\right)
  \left[(1+{\rm i}{\rm We})\mu+
  (\mu_{\rm b}-{\textstyle{{2}\over{3}}}\mu)\right]\right\}\nonumber\\
  &&+{{{\rm i}\Sigma}\over{\Omega}}
  \left({{{\rm d}p}\over{{\rm d}\Sigma}}\right).
\end{eqnarray}
Instability on small scales occurs if ${\rm Re}({\sA})<0$, and one finds
\begin{eqnarray}
  \lefteqn{(1+{\rm We}^2)^2{\rm Re}({\sA})=
  3(1+4{\rm We}^2-{\rm We}^4)\mu}&\nonumber\\
  &&-3(1+{\rm We}^2)\Sigma\left({{{\rm d}\mu}\over{{\rm d}\Sigma}}\right)+
  (1+7{\rm We}^2)(\mu_{\rm b}-{\textstyle{{2}\over{3}}}\mu).
\end{eqnarray}
For either Thomson or Kramers opacity, the instability persists for
all values of ${\rm We}$ if $\mu_{\rm b}/\mu=0$.  However, for
reasonable values of the bulk viscosity and relaxation time, the
instability can be quenched.  For example, the case $\mu_{\rm
  b}/\mu=2/3$ corresponds to the vanishing of the `bulk term' in
equation (\ref{stress}).  For this value, the instability is quenched
for $3^{-1/2}<{\rm We}<2^{1/2}$ (i.e. $0.577<{\rm We}<1.414$) in the
case of Thomson opacity, and for $0.423<{\rm We}<1.547$ in the case of
Kramers opacity.  It is tantalizing that the value tentatively deduced
from simulations of MHD turbulence, ${\rm We}\approx0.5$, may or may
not be sufficient to stabilize the disc, and the outcome may differ
according to the opacity law.

A related result is that the viscous overstability of axisymmetric
waves can be quenched by introducing a more sophisticated turbulence
model with a non-zero relaxation time (Kato 1994).  However, a direct
comparison with Kato's work is not possible.  His second-order closure
model is aimed at modelling the turbulent Reynolds stress in an
incompressible fluid that undergoes a notional hydrodynamic
instability, whereas the Maxwellian viscoelastic model used in the
present paper is aimed at modelling (predominantly) the turbulent
Maxwell stress in a compressible fluid that undergoes the
magnetorotational instability.

\section{THE FULL MODEL}

\subsection{Basic equations}

\label{basic}

In this section the basic equations governing a fluid disc in three
dimensions are expressed in a general, time-independent coordinate
system.  Following the usual notation of tensor calculus, $g_{ab}$
denotes the metric tensor and $\nabla_a$ the covariant derivative.

The equation of mass conservation is
\begin{equation}
  (\partial_t+u^a\nabla_a)\rho=-\rho\nabla_au^a,
  \label{drho}
\end{equation}
where $\rho$ is the density and $\bu$ the velocity.  The equation of
motion is
\begin{equation}
  \rho(\partial_t+u^b\nabla_b)u^a=-\rho\nabla^a\Phi-\nabla^ap+
  \nabla_bT^{ab}+\rho f^a,
  \label{du}
\end{equation}
where $\Phi$ is the gravitational potential, $p$ the pressure, ${\bf
  T}$ the turbulent stress tensor and $\bff$ the external force per
unit mass.

Under the assumption that the turbulent stress behaves similarly to a
Maxwellian viscoelastic fluid, the stress tensor satisfies equation
(\ref{stress}), i.e.
\begin{eqnarray}
  \lefteqn{T^{ab}+\tau\left[(\partial_t+u^c\nabla_c)T^{ab}-
  T^{ac}\nabla_cu^b-T^{bc}\nabla_cu^a+
  2(\nabla_cu^c)T^{ab}\right]}&\nonumber\\
  &&=\mu(\nabla^au^b+\nabla^bu^a)+
  (\mu_{\rm b}-{\textstyle{{2}\over{3}}}\mu)(\nabla_cu^c)g^{ab},
  \label{tab}
\end{eqnarray}
where $\tau$ is the relaxation time, $\mu$ the effective shear
viscosity and $\mu_{\rm b}$ the effective bulk viscosity.  The energy
equation is
\begin{eqnarray}
  \lefteqn{\left({{1}\over{\gamma-1}}\right)(\partial_t+u^a\nabla_a)p=
  -\left({{\gamma}\over{\gamma-1}}\right)p\nabla_au^a}&\nonumber\\
  &&+T^{ab}\nabla_au_b-\nabla_aF^a_{\rm rad},
  \label{dp}
\end{eqnarray}
where $\gamma$ is the adiabatic exponent and $\bF_{\rm rad}$ the
radiative energy flux, given in the Rosseland approximation for an
optically thick medium by
\begin{equation}
  F^a_{\rm rad}=-{{16\sigma T^3}\over{3\kappa\rho}}\nabla^aT,
\end{equation}
where $\sigma$ is the Stefan-Boltzmann constant, $T$ the temperature
and $\kappa$ the opacity.  The equation of state of an ideal gas,
\begin{equation}
  p={{k\rho T}\over{\mu_{\rm m}m_{\rm H}}},
\end{equation}
is adopted, where $k$ is Boltzmann's constant, $\mu_{\rm m}$ the mean
molecular weight and $m_{\rm H}$ the mass of the hydrogen atom.  The
opacity is assumed to be of the power-law form
\begin{equation}
  \kappa=C_\kappa\rho^xT^y,
\end{equation}
where $C_\kappa$ is a constant.  This includes the important cases of
Thomson scattering opacity ($x=y=0$) and Kramers opacity ($x=1$,
$y=-7/2$).

The gravitational potential is $\Phi=-GM/r$, where $r$ is the distance
from the central object.  The self-gravitation of the disc is
neglected.  The effective viscosity coefficients are assumed to be
given by an alpha parametrization.  The precise form of the alpha
prescription relevant to an eccentric disc is to some extent
debatable.  It will be convenient to adopt the form
\begin{equation}
  \mu=\alpha p\left({{GM}\over{\lambda^3}}\right)^{-1/2},\qquad
  \mu_{\rm b}=\alpha_{\rm b}p\left({{GM}\over{\lambda^3}}\right)^{-1/2},
\end{equation}
where $\alpha$ and $\alpha_{\rm b}$ are the dimensionless shear and
bulk viscosity parameters, and $\lambda$ is the semi-latus rectum of
the elliptical orbits (introduced below).  In the limit of a circular
disc, this prescription reduces to the usual one, $\mu=\alpha
p/\Omega$, etc., where $\Omega$ is the orbital angular velocity.
Finally, the Weissenberg number is defined by
\begin{equation}
  {\rm We}=\tau\left({{GM}\over{\lambda^3}}\right)^{1/2},
\end{equation}
where $\tau$ is the relaxation time.  Again, this reduces to ${\rm
  We}=\Omega\tau$ in the limit of a circular disc.

\subsection{Orbital coordinates}

Consider a thin, eccentric accretion disc in which the dominant
motion consists of coplanar, elliptical Keplerian orbits.  Let
$(R,\phi,z)$ be cylindrical polar coordinates such that the mid-plane
of the disc corresponds to $z=0$.  Then the shape of the orbits is
given by
\begin{equation}
  R=\lambda(1+e\cos\theta)^{-1},
  \label{r}
\end{equation}
where $\lambda$ is the semi-latus rectum, $e$ the eccentricity and
$\theta=\phi-\omega$ the azimuth relative to periastron (known as the
`true anomaly' in celestial mechanics).  The semi-latus rectum is
related to the specific angular momentum $h=(GM\lambda)^{1/2}$ and is
conveniently adopted as a label for the different orbits.  In general
the eccentricity and longitude of periastron may vary continuously
from one orbit to the next and also with time, so that
$e=e(\lambda,t)$ and $\omega=\omega(\lambda,t)$.  This is illustrated
in Fig.~1.  Let $\dot e$ and $e'$ then denote $\partial_t e$ and
$\partial_\lambda e$, etc.  It will also be convenient to define the
complex eccentricity $E=e\,{\rm e}^{{\rm i}\omega}$.

\begin{figure}
  \centerline{\epsfbox{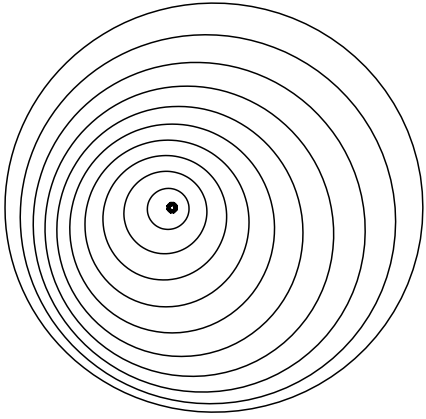}}
  \caption{Example of instantaneous orbits in an eccentric disc.}
\end{figure}

The analysis of an eccentric disc is greatly facilitated by the
introduction of {\it orbital coordinates\/} $(\lambda,\phi)$ in place of
$(R,\phi)$.  However, some preliminary analysis must be carried out
because these coordinates are both non-orthogonal and time-dependent.

\subsubsection{Spatial derivatives}

To evaluate the spatial derivatives in the governing equations, one
must obtain the metric coefficients and connection components.  The
two-dimensional line element is given by
\begin{equation}
  {\rm d}s^2={\rm d}R^2+R^2\,{\rm d}\phi^2=
  (R_\lambda\,{\rm d}\lambda+R_\phi\,{\rm d}\phi)^2+R^2\,{\rm d}\phi^2,
\end{equation}
where the subscripts on $R$ stand for partial derivatives of the
function $R(\lambda,\phi,t)$.  These can be evaluated from equation
(\ref{r}), but the more general notation will be retained.  Thus the
metric coefficients are
\begin{equation}
  g_{\lambda\lambda}=R_\lambda^2,\qquad
  g_{\lambda\phi}=R_\lambda R_\phi,\qquad
  g_{\phi\phi}=R^2+R_\phi^2.
\end{equation}
As expected, the square root of the metric determinant is equal to the
Jacobian of the coordinate system:
\begin{equation}
  g^{1/2}=J={{\partial(x,y)}\over{\partial(\lambda,\phi)}}
  ={{\partial(x,y)}\over{\partial(R,\phi)}}
  {{\partial(R,\phi)}\over{\partial(\lambda,\phi)}}=RR_\lambda.
\end{equation}
The inverse metric coefficients are
\begin{equation}
  g^{\lambda\lambda}={{R^2+R_\phi^2}\over{R^2R_\lambda^2}},\qquad
  g^{\lambda\phi}=-{{R_\phi}\over{R^2R_\lambda}},\qquad
  g^{\phi\phi}={{1}\over{R^2}}.
\end{equation}
The components of the metric connection, given by
\begin{equation}
  \Gamma^a_{bc}={\textstyle{{1}\over{2}}}g^{ad}
  (\partial_b g_{cd}+\partial_c g_{db}-\partial_d g_{bc}),
\end{equation}
are found to be
\begin{equation}
  \Gamma^\lambda_{\lambda\lambda}=
  {{R_{\lambda\lambda}}\over{R_\lambda}},\qquad
  \Gamma^\lambda_{\lambda\phi}={{R_{\lambda\phi}}\over{R_\lambda}}-
  {{R_\phi}\over{R}},
\end{equation}
\begin{equation}
  \Gamma^\lambda_{\phi\phi}=-{{(R^2+2R_\phi^2-RR_{\phi\phi})}\over
  {RR_\lambda}},
\end{equation}
\begin{equation}
  \Gamma^\phi_{\lambda\lambda}=0,\qquad
  \Gamma^\phi_{\lambda\phi}={{R_\lambda}\over{R}},\qquad
  \Gamma^\phi_{\phi\phi}={{2R_\phi}\over{R}}.
\end{equation}

The coordinate system is trivially extended to three dimensions by
incorporating the vertical coordinate $z$.  With the exception of
$g_{zz}=g^{zz}=1$, all metric coefficients and connection components
involving $z$ vanish.  The divergences of a vector and of a symmetric
tensor in three dimensions are then
\begin{eqnarray}
  \nabla_au^a&=&\partial_au^a+\Gamma^a_{ba}u^b,\\
  \nabla_bT^{ab}&=&\partial_bT^{ab}+\Gamma^a_{cb}T^{cb}+\Gamma^b_{cb}T^{ac}.
\end{eqnarray}
The vector divergence can be written explicitly as
\begin{equation}
  {{1}\over{J}}\partial_\lambda(Ju^\lambda)+
  {{1}\over{J}}\partial_\phi(Ju^\phi)+\partial_zu^z,
\end{equation}
while the $\lambda$-, $\phi$- and $z$-components of the tensor divergence are
\begin{eqnarray}
  \lefteqn{{{1}\over{JR_\lambda}}\partial_\lambda
  (JR_\lambda T^{\lambda\lambda})+
  {{R^2}\over{JR_\lambda^2}}\partial_\phi
  \left({{JR_\lambda^2}\over{R^2}}T^{\lambda\phi}\right)}&\nonumber\\
  &&-{{1}\over{RR_\lambda}}(R^2+2R_\phi^2-RR_{\phi\phi})T^{\phi\phi}+
  \partial_zT^{\lambda z},
\end{eqnarray}
\begin{equation}
  {{1}\over{JR^2}}\partial_\lambda(JR^2T^{\lambda\phi})+
  {{1}\over{JR^2}}\partial_\phi(JR^2T^{\phi\phi})+\partial_zT^{\phi z},
\end{equation}
\begin{equation}
  {{1}\over{J}}\partial_\lambda(JT^{\lambda z})+
  {{1}\over{J}}\partial_\phi(JT^{\phi z})+\partial_zT^{zz},
\end{equation}
respectively.

For a physically meaningful solution, the eccentricity is restricted
in magnitude.  In order that the orbits be closed, one obviously
requires $|E|<1$.  A further restriction is that the Jacobian
determinant should remain positive, which implies $|E-\lambda E'|<1$.
If this condition were violated, neighbouring orbits would intersect
each other.

\subsubsection{Time-derivatives}

The equations of Section~\ref{basic} are valid only in a time-independent
(`inertial') coordinate system.  To adapt them to orbital coordinates
$q^a=(\lambda,\phi,z)$, observe that the partial derivatives at fixed
inertial and orbital coordinates are related by
\begin{equation}
  (\partial_t)_{\rm inertial}=(\partial_t)_{\rm orbital}+
  \dot q^a\partial_a,
\end{equation}
where
\begin{equation}
  \dot q^a=(\partial_t)_{\rm inertial}q^a.
\end{equation}
Thus, for a scalar quantity such as
$\rho$ in equation (\ref{drho}) or $p$ in equation (\ref{dp}), one
replaces
\begin{equation}
  \partial_t\rho\mapsto\partial_t\rho+\dot q^a\partial_a\rho.
\end{equation}
When differentiating the vector $\bu$ in equation (\ref{du}) one must
also take into account the time-dependence of the orbital basis
vectors.  One should then replace
\begin{equation}
  \partial_tu^a\mapsto\partial_tu^a+\dot q^b\partial_bu^a-
  u^b\partial_b\dot q^a.
\end{equation}
This is most easily derived by noting that
$u^\lambda=\bu\!\cdot\!\nabla\lambda$, etc., and obtaining the
equation for the time-derivative of this quantity.  More generally,
one is replacing $\partial_t$ with $\partial_t+\pounds_\bqdot$, where
$\pounds$ denotes the Lie derivative.  Thus in equation (\ref{tab})
one should replace
\begin{equation}
  \partial_tT^{ab}\mapsto\partial_tT^{ab}+\dot q^c\partial_cT^{ab}-
  T^{cb}\partial_c\dot q^a-T^{ac}\partial_c\dot q^b.
\end{equation}

Hereafter $\partial_t$ will represent the derivative at fixed orbital
coordinates.  For orbital coordinates only $\dot\lambda$ is non-zero,
so $\dot q^a\partial_a=\dot\lambda\partial_\lambda$.  The identity
\begin{equation}
  \partial_t J+\partial_\lambda(J\dot\lambda)
  \label{dj}
\end{equation}
will be used below.  This is established by noting that
$(\partial_t)_{\rm inertial}R=0$, so that
\begin{equation}
  \partial_t R+\dot\lambda R_\lambda=0,
\end{equation}
and so
\begin{equation}
  \partial_t J+\partial_\lambda(J\dot\lambda)=
  \partial_t(RR_\lambda)-\partial_\lambda(R\partial_t R)=0,
\end{equation}
as required.

\section{Asymptotic development}

\subsection{Scaled variables and expansions}

The key to analysing this problem, as with many dynamical problems in
accretion discs, is to recognize that there is a separation of scales.
The disc is thin, in the sense that the semi-thickness $H$ of the disc
satisfies $H/R\ll1$.  Additionally, the evolution of surface density
and eccentricity occurs on a time-scale much longer than the orbital
time-scale of the fluid.  These ideas lead naturally to the
introduction of scaled variables and an asymptotic development.

Let the small parameter $\epsilon$ be a characteristic value of the
angular semi-thickness $H/R$ of the disc.  Adopt a system of units
such that the radius of the disc and the characteristic orbital
time-scale are $O(1)$.  Then define the stretched vertical coordinate
$\zeta=z/\epsilon$, which is $O(1)$ inside the disc.  The slow
evolution is captured by a slow time coordinate $T=\epsilon^2t$ and
one may replace
\begin{equation}
  e(\lambda,t)\mapsto e(\lambda,T),\qquad
  \omega(\lambda,t)\mapsto\omega(\lambda,T).
\end{equation}

For the fluid variables, introduce the expansions
\begin{eqnarray}
  u^\lambda&=&\epsilon^2u^\lambda_2(\lambda,\phi,\zeta,T)+
  O(\epsilon^4),\\
  u^\phi&=&\Omega(\lambda,\phi,T)+
  \epsilon^2u^\phi_2(\lambda,\phi,\zeta,T)+O(\epsilon^4),\\
  u^z&=&\epsilon u^z_1(\lambda,\phi,\zeta,T)+
  \epsilon^3u^z_3(\lambda,\phi,\zeta,T)+O(\epsilon^5),\\
  \rho&=&\epsilon^s\left[\rho_0(\lambda,\phi,\zeta,T)+
  \epsilon^2\rho_2(\lambda,\phi,\zeta,T)+O(\epsilon^4)\right],\\
  p&=&\epsilon^{s+2}\left[p_0(\lambda,\phi,\zeta,T)+O(\epsilon^2)\right],\\
  T^{\lambda\lambda}&=&\epsilon^{s+2}\left[
  T^{\lambda\lambda}_0(\lambda,\phi,\zeta,T)+O(\epsilon^2)\right],\\
  T^{\lambda\phi}&=&\epsilon^{s+2}\left[
  T^{\lambda\phi}_0(\lambda,\phi,\zeta,T)+O(\epsilon^2)\right],\\
  T^{\phi\phi}&=&\epsilon^{s+2}\left[
  T^{\phi\phi}_0(\lambda,\phi,\zeta,T)+O(\epsilon^2)\right],\\
  T^{\lambda z}&=&\epsilon^{s+3}\left[
  T^{\lambda z}_1(\lambda,\phi,\zeta,T)+O(\epsilon^2)\right],\\
  T^{\phi z}&=&\epsilon^{s+3}\left[
  T^{\phi z}_1(\lambda,\phi,\zeta,T)+O(\epsilon^2)\right],\\
  T^{zz}&=&\epsilon^{s+2}\left[
  T^{zz}_0(\lambda,\phi,\zeta,T)+O(\epsilon^2)\right],\\
  \mu&=&\epsilon^{s+2}\left[\mu_0(\lambda,\phi,\zeta,T)+
  O(\epsilon^2)\right],\\
  \mu_{\rm b}&=&\epsilon^{s+2}\left[\mu_{{\rm b}0}(\lambda,\phi,\zeta,T)+
  O(\epsilon^2)\right],\\
  T&=&\epsilon^2\left[T_0(\lambda,\phi,\zeta,T)+O(\epsilon^2)\right],\\
  F^z_{\rm rad}&=&\epsilon^{s+3}\left[F_0(\lambda,\phi,\zeta,T)+
  O(\epsilon^2)\right],
\end{eqnarray}
while the horizontal components of $\bF_{\rm rad}$ are
$O(\epsilon^{s+4})$.  Here $s$ is a positive parameter, which drops
out of the analysis, although formally one requires $s=(4-2y)/(2+x)$
in order to balance powers of $\epsilon$ in the opacity law.  Note
that the dominant motion is an orbital motion with angular velocity
$\Omega$, independent of $\zeta$.

The potential is expanded in a Taylor series,
\begin{equation}
  \Phi=\Phi_0(\lambda,\phi,T)+{\textstyle{{1}\over{2}}}\epsilon^2\zeta^2
  \Phi_2(\lambda,\phi,T)+O(\epsilon^4),
\end{equation}
where
\begin{equation}
  \Phi_0=-{{GM}\over{R}},\qquad
  \Phi_2={{GM}\over{R^3}}.
\end{equation}
The external force per unit mass is also expanded as
\begin{eqnarray}
  f^\lambda&=&\epsilon^2\left[f^\lambda_0(\lambda,\phi,T)+
  O(\epsilon^2)\right],\\
  f^\phi&=&\epsilon^2\left[f^\phi_0(\lambda,\phi,T)+
  O(\epsilon^2)\right],\\
  f^z&=&O(\epsilon^3).
\end{eqnarray}
This is the correct form if $\bff$ represents the tidal force due to a
companion object in a coplanar orbit.  The $\epsilon^2$ scaling will
be seen to be appropriate, and indicates that the external force is
small compared to the gravitational force of the central object.

When these expansions are substituted into the equations of Section~2,
various equations are obtained at different orders in $\epsilon$.  The
required equations comprise three sets, representing three different
physical problems, and these will be considered in turn.

\subsection{Orbital motion}

The horizontal components of the equation of motion (\ref{du}) at
leading order [$O(\epsilon^s)$] are
\begin{equation}
  \rho_0\Gamma^\lambda_{\phi\phi}\Omega^2=
  -\rho_0(g^{\lambda\lambda}\partial_\lambda\Phi_0+
  g^{\lambda\phi}\partial_\phi\Phi_0),
\end{equation}
\begin{equation}
  \rho_0(\Omega\partial_\phi\Omega+\Gamma^\phi_{\phi\phi}\Omega^2)=
  -\rho_0(g^{\lambda\phi}\partial_\lambda\Phi_0+
  g^{\phi\phi}\partial_\phi\Phi_0).
\end{equation}
Since $\Phi_0$ is a function of $R$ only, these simplify to
\begin{equation}
  (R^2+2R_\phi^2-RR_{\phi\phi})\Omega^2=R{{{\rm d}\Phi_0}\over{{\rm d}R}},
\end{equation}
\begin{equation}
  \partial_\phi(R^2\Omega)=0,
  \label{dphir2om}
\end{equation}
and are satisfied by the angular velocity of an elliptical Keplerian orbit,
\begin{equation}
  \Omega=\left({{GM}\over{\lambda^3}}\right)^{1/2}(1+e\cos\theta)^2,
\end{equation}
where $e(\lambda,T)$ and $\omega(\lambda,T)$ are arbitrary.  It
is useful to note from the above that
\begin{equation}
  R^2+2R_\phi^2-RR_{\phi\phi}={{R^3}\over{\lambda}},
  \label{useful}
\end{equation}
which simplifies the expression for one of the connection components,
\begin{equation}
  \Gamma^\lambda_{\phi\phi}=-{{R^2}\over{\lambda R_\lambda}}.
\end{equation}
The orbital period is\footnote{Throughout, integrals with respect to
  $\phi$ are carried out from $0$ to $2\pi$, and integrals with
  respect to $\zeta$ over the full vertical extent of the disc.}
\begin{equation}
  P=\int\,{{{\rm d}\phi}\over{\Omega}}=
  2\pi(1-e^2)^{-3/2}\left({{\lambda^3}\over{GM}}\right)^{1/2}.
\end{equation}

\subsection{Slow velocities and time-evolution}

The equation of mass conservation (\ref{drho}) at leading order
[$O(\epsilon^s)$] is
\begin{equation}
  (\Omega\partial_\phi+u^z_1\partial_\zeta)\rho_0=
  -\rho_0\left[{{1}\over{J}}\partial_\phi(J\Omega)+\partial_\zeta u^z_1\right].
  \label{rho0a}
\end{equation}
Introduce the surface density at leading order [$O(\epsilon^{s+1})$],
\begin{equation}
  \tilde\Sigma(\lambda,\phi,T)=\int\rho_0\,{\rm d}\zeta.
\end{equation}
Then the vertically integrated version of equation (\ref{rho0a}) may
be written in the form
\begin{equation}
  \partial_\phi(J\tilde\Sigma\Omega)=0,
  \label{jsigom}
\end{equation}
which determines the variation of surface density around the orbit due
to eccentricity.  It will be convenient to introduce a pseudo-circular
surface density,
\begin{equation}
  \Sigma(\lambda,T)={{1}\over{2\pi\lambda}}
  \int J\tilde\Sigma\,{\rm d}\phi,
\end{equation}
which has the property that the mass contained between orbits
$\lambda_1$ and $\lambda_2$ is
\begin{equation}
  2\pi\int_{\lambda_1}^{\lambda_2}\Sigma\,\lambda\,{\rm d}\lambda.
\end{equation}
It will also be useful to introduce the second vertical moment of the
density,
\begin{equation}
  \tilde{\sI}(\lambda,\phi,T)=\int\rho_0\zeta^2\,{\rm d}\zeta,
\end{equation}
and the pseudo-circular average,
\begin{equation}
  {\sI}(\lambda,T)={{1}\over{2\pi\lambda}}
  \int J\tilde{\sI}\,{\rm d}\phi.
\end{equation}

The equation of mass conservation (\ref{drho}) at $O(\epsilon^{s+2})$ is
\begin{eqnarray}
  \lefteqn{\left[\partial_T+(u^\lambda_2+\dot\lambda)\partial_\lambda+
  u^\phi_2\partial_\phi+u^z_3\partial_\zeta\right]\rho_0+
  (\Omega\partial_\phi+u^z_1\partial_\zeta)\rho_2}&\nonumber\\
  &&=-\rho_0\left[{{1}\over{J}}\partial_\lambda(Ju^\lambda_2)+
  {{1}\over{J}}\partial_\phi(Ju^\phi_2)+\partial_\zeta u^z_3\right]\nonumber\\
  &&-\rho_2\left[{{1}\over{J}}\partial_\phi(J\Omega)+
  \partial_\zeta u^z_1\right].
  \label{rho2}
\end{eqnarray}
The horizontal components of the equation of motion (\ref{du}) at
$O(\epsilon^{s+2})$ are
\begin{eqnarray}
  \lefteqn{\rho_0\left[(\Omega\partial_\phi+u^z_1\partial_\zeta)u^\lambda_2-
  \Omega\partial_\phi\dot\lambda+
  2\Gamma^\lambda_{\lambda\phi}\Omega u^\lambda_2+
  2\Gamma^\lambda_{\phi\phi}\Omega u^\phi_2\right]}&\nonumber\\
  &&=-{{\rho_0\zeta^2}\over{2R_\lambda}}{{{\rm d}\Phi_2}\over{{\rm d}R}}-
  g^{\lambda\lambda}\partial_\lambda p_0-
  g^{\lambda\phi}\partial_\phi p_0\nonumber\\
  &&+{{1}\over{JR_\lambda}}\partial_\lambda(JR_\lambda T^{\lambda\lambda}_0)+
  {{R^2}\over{JR_\lambda^2}}\partial_\phi
  \left({{JR_\lambda^2}\over{R^2}}T^{\lambda\phi}_0\right)\nonumber\\
  &&-{{R^2}\over{\lambda R_\lambda}}T^{\phi\phi}_0+
  \partial_\zeta T^{\lambda z}_1+\rho_0f^\lambda_0,
  \label{ulambda2}
\end{eqnarray}
\begin{eqnarray}
  \lefteqn{\rho_0\left[\partial_T+(u^\lambda_2+\dot\lambda)\partial_\lambda+
  u^\phi_2\partial_\phi\right]\Omega}&\nonumber\\
  &&+\rho_0\left[(\Omega\partial_\phi+u^z_1\partial_\zeta)u^\phi_2+
  2\Gamma^\phi_{\lambda\phi}\Omega u^\lambda_2+
  2\Gamma^\phi_{\phi\phi}\Omega u^\phi_2\right]\nonumber\\
  &&=-g^{\lambda\phi}\partial_\lambda p_0-g^{\phi\phi}\partial_\phi p_0+
  {{1}\over{JR^2}}\partial_\lambda(JR^2T^{\lambda\phi}_0)\nonumber\\
  &&+{{1}\over{JR^2}}\partial_\phi(JR^2T^{\phi\phi}_0)+
  \partial_\zeta T^{\phi z}_1+\rho_0f^\phi_0.
  \label{uphi2}
\end{eqnarray}
It will be sufficient to work with the vertically integrated versions
of equations (\ref{rho2}), (\ref{ulambda2}) and (\ref{uphi2}), which
may be written in the form
\begin{eqnarray}
  \lefteqn{(\partial_T+\dot\lambda\partial_\lambda)\tilde\Sigma+
  {{1}\over{J}}\partial_\lambda\int J\rho_0u^\lambda_2\,{\rm d}\zeta}
  &\nonumber\\
  &&+{{1}\over{J}}\partial_\phi\int J(\rho_0u^\phi_2+\rho_2\Omega)
  \,{\rm d}\zeta=0,
  \label{dsigma}
\end{eqnarray}
\begin{equation}
  \Omega\partial_\phi(v^\lambda-\dot\lambda)+
  2\Gamma^\lambda_{\lambda\phi}\Omega v^\lambda+
  2\Gamma^\lambda_{\phi\phi}\Omega v^\phi=A^\lambda,
  \label{vlambda}
\end{equation}
\begin{eqnarray}
  \lefteqn{\left[\partial_T+(v^\lambda+\dot\lambda)\partial_\lambda+
  v^\phi\partial_\phi\right]\Omega+\Omega\partial_\phi v^\phi}&\nonumber\\
  &&+2\Gamma^\phi_{\lambda\phi}\Omega v^\lambda+
  2\Gamma^\phi_{\phi\phi}\Omega v^\phi=A^\phi,
  \label{vphi}
\end{eqnarray}
where $\bv(\lambda,\phi,T)$ is a mean horizontal velocity defined
by
\begin{equation}
  \tilde\Sigma v^a=\int\rho_0u^a_2\,{\rm d}\zeta,
\end{equation}
and $\bA(\lambda,\phi,T)$ a mean horizontal acceleration given by
\begin{eqnarray}
  \lefteqn{\tilde\Sigma A^\lambda=-{{\tilde{\sI}}\over{2R_\lambda}}
  {{{\rm d}\Phi_2}\over{{\rm d}R}}+
  {{1}\over{JR_\lambda}}\partial_\lambda
  (JR_\lambda{\sT}^{\lambda\lambda})}&\nonumber\\
  &&+{{R^2}\over{JR_\lambda^2}}\partial_\phi
  \left({{JR_\lambda^2}\over{R^2}}{\sT}^{\lambda\phi}\right)-
  {{R^2}\over{\lambda R_\lambda}}{\sT}^{\phi\phi}+
  \tilde\Sigma f^\lambda_0,
\end{eqnarray}
\begin{equation}
  \tilde\Sigma A^\phi={{1}\over{JR^2}}\partial_\lambda
  (JR^2{\sT}^{\lambda\phi})+
  {{1}\over{JR^2}}\partial_\phi(JR^2{\sT}^{\phi\phi})+
  \tilde\Sigma f^\phi_0,
\end{equation}
where
\begin{equation}
  {\sT}^{ab}=\int(T^{ab}_0-p_0g^{ab})\,{\rm d}\zeta
\end{equation}
is the vertically integrated stress tensor including the pressure term.

The objective at this stage is to extract evolutionary equations for
$\Sigma$ and $E$ without attempting to obtain a complete solution for
$v^\lambda$ and $v^\phi$.  First, equation (\ref{dsigma}) is
integrated over $\phi$ to eliminate the mass flux around the orbit and
obtain
\begin{equation}
  \partial_T\int J\tilde\Sigma\,{\rm d}\phi+
  \partial_\lambda\int J\tilde\Sigma(v^\lambda+\dot\lambda)\,{\rm d}\phi=0,
  \label{dintsigma}
\end{equation}
where the identity (\ref{dj}) has been used.  This expresses the
conservation of mass in one dimension and involves the {\it relative
  velocity\/} $(v^\lambda+\dot\lambda)$.  Let $v(\lambda,T)$ be a mean
accretion velocity defined by
\begin{equation}
  v\int J\tilde\Sigma\,{\rm d}\phi=\int J\tilde\Sigma
  (v^\lambda+\dot\lambda)\,{\rm d}\phi.
\end{equation}
Then equation (\ref{dintsigma}) may be written in the form
\begin{equation}
  \dot\Sigma+{{1}\over{\lambda}}\partial_\lambda(\lambda v\Sigma)=0,
\end{equation}
formally identical to the equation of mass conservation for a
circular disc.

Next, equations (\ref{vlambda}) and (\ref{vphi}) may be rewritten in
the form
\begin{equation}
  \Omega\partial_\phi\left({{R_\lambda^2}\over{R^2}}v^\lambda\right)-
  {{2R_\lambda}\over{\lambda}}\Omega v^\phi=
  {{R_\lambda^2}\over{R^2}}B^\lambda,
  \label{vlambda2}
\end{equation}
\begin{equation}
  \Omega\partial_\phi(R^2v^\phi)+v^\lambda{{{\rm d}h}\over{{\rm d}\lambda}}
  =R^2B^\phi,
  \label{vphi2}
\end{equation}
where $h=(GM\lambda)^{1/2}$ is the specific angular momentum and
\begin{equation}
  B^\lambda=A^\lambda+\Omega\partial_\phi\dot\lambda,\qquad
  B^\phi=A^\phi-{{\dot\lambda}\over{R^2}}{{{\rm d}h}\over{{\rm d}\lambda}}.
\end{equation}
Equation (\ref{vphi2}) expresses the conservation of angular momentum
and may be integrated over the orbit to obtain
\begin{equation}
  {{{\rm d}h}\over{{\rm d}\lambda}}\int J\tilde\Sigma
  (v^\lambda+\dot\lambda)\,{\rm d}\phi=
  \int J\tilde\Sigma R^2A^\phi\,{\rm d}\phi,
\end{equation}
which involves exactly the same average relative velocity as in
equation (\ref{dintsigma}).  Indeed, these two equations clearly
relate the evolution of mass to the total torque, as expected on
general grounds.  This may be simplified to
\begin{equation}
  v{{{\rm d}h}\over{{\rm d}\lambda}}={{1}\over{P}}\int R^2A^\phi
  \,{{{\rm d}\phi}\over{\Omega}}.
  \label{compare1}
\end{equation}
Using the expression for $A^\phi$, this becomes
\begin{equation}
  v{{{\rm d}h}\over{{\rm d}\lambda}}\int J\tilde\Sigma\,{\rm d}\phi=
  \partial_\lambda\int JR^2{\sT}^{\lambda\phi}\,{\rm d}\phi+
  \int J\tilde\Sigma R^2f^\phi_0\,{\rm d}\phi.
\end{equation}

It remains to determine the evolution of the eccentricity.  Here one
uses the fact that the operator acting on the unknown $\bv$ in
equations (\ref{vlambda2}) and (\ref{vphi2}) is singular, and the
equations can therefore be solved (in principle) only when $\bB$
satisfies a certain solvability condition.  Eliminate $v^\lambda$
between equations (\ref{vlambda2}) and (\ref{vphi2}) to obtain
\begin{equation}
  {\sL}v^\phi={\sR},
  \label{lvphi}
\end{equation}
where the linear operator ${\sL}$ is defined by
\begin{equation}
  {\sL}v^\phi=\Omega\partial_\phi\left[{{R_\lambda^2}\over{R^2}}
  \Omega\partial_\phi(R^2v^\phi)\right]+
  2{{{\rm d}h}\over{{\rm d}\lambda}}{{R_\lambda}\over{\lambda}}\Omega v^\phi,
\end{equation}
and the right-hand side is
\begin{equation}
  {\sR}=\Omega\partial_\phi(R_\lambda^2B^\phi)-
  {{{\rm d}h}\over{{\rm d}\lambda}}{{R_\lambda^2}\over{R^2}}B^\lambda.
\end{equation}
Now ${\sL}$ is a singular operator because it possesses a null
eigenvector
\begin{equation}
  w^\phi={{{\rm e}^{{\rm i}\phi}}\over{R_\lambda}}
\end{equation}
such that ${\sL}w^\phi=0$.  This can be verified by direct
substitution, with the help of equations (\ref{dphir2om}) and
(\ref{useful}).  This null solution corresponds simply to a small
redefinition of the eccentricity, $E\mapsto E+\delta E$.  The
corresponding solvability condition for equation (\ref{lvphi}) is
\begin{equation}
  \int J\tilde\Sigma R^2w^\phi{\sR}\,{\rm d}\phi=0.
\end{equation}
Multiplying this by $2{\rm i}\lambda/GM$ and taking into account the
contributions to $\bB$ from $\dot\lambda$, this can be brought into
the form
\begin{eqnarray}
  \lefteqn{\dot E\int J\tilde\Sigma\,{\rm d}\phi=}&\nonumber\\
  &&{{2{\rm i}\lambda}\over{GM}}
  \int J\tilde\Sigma{{R^2{\rm e}^{{\rm i}\phi}}\over{R_\lambda}}
  \left[\Omega\partial_\phi(R_\lambda^2A^\phi)-
  {{{\rm d}h}\over{{\rm d}\lambda}}{{R_\lambda^2}\over{R^2}}A^\lambda\right]
  \,{\rm d}\phi,
\end{eqnarray}
where the relation
\begin{equation}
  {{R_\lambda\dot\lambda}\over{R}}={{\dot e\cos\theta+e\dot\omega\sin\theta}
  \over{1+e\cos\theta}}
\end{equation}
has been used, and also the integrals
\begin{equation}
  \int_0^{2\pi}(1+e\cos\theta)^{-3}\,{\rm d}\theta=
  (2+e^2)\pi(1-e^2)^{-5/2},
\end{equation}
\begin{equation}
  \int_0^{2\pi}(1+e\cos\theta)^{-3}\cos\theta\,{\rm d}\theta=
  -3e\pi(1-e^2)^{-5/2},
\end{equation}
\begin{equation}
  \int_0^{2\pi}(1+e\cos\theta)^{-3}\cos^2\theta\,{\rm d}\theta=
  (1+2e^2)\pi(1-e^2)^{-5/2},
\end{equation}
which are easily established by the residue theorem.  Thus the
required equation for the evolution of $E$ has been obtained, but must
now be manipulated into a more usable form.

With the help of the relation
\begin{equation}
  {{\lambda^2R_\lambda^2}\over{R^4}}\partial_\phi
  \left({{R^2{\rm e}^{{\rm i}\phi}}\over{R_\lambda}}\right)=
  {\rm i}({\rm e}^{{\rm i}\phi}+E-\lambda E'),
\end{equation}
one obtains
\begin{eqnarray}
  \lefteqn{\dot E\int J\tilde\Sigma\,{\rm d}\phi
  =-{{{\rm i}}\over{GM}}\int J\tilde\Sigma\Omega R^2R_\lambda
  {\rm e}^{{\rm i}\phi}A^\lambda\,{\rm d}\phi}&\nonumber\\
  &&+{{2}\over{GM\lambda}}
  \int J\tilde\Sigma\Omega({\rm e}^{{\rm i}\phi}+E-\lambda E')
  R^4A^\phi\,{\rm d}\phi.
\end{eqnarray}
This can be further simplified to
\begin{equation}
  h\left(\dot E+vE'-{{vE}\over{\lambda}}\right)=
  {{1}\over{P}}\int(2R^2A^\phi-{\rm i}\lambda R_\lambda A^\lambda)
  \,{{\rm e}^{{\rm i}\phi}}\,{{{\rm d}\phi}\over{\Omega}}.
  \label{compare2}
\end{equation}
Taking into account the various contributions to $\bA$, this becomes
\begin{eqnarray}
  h\lefteqn{\left(\dot E+vE'-{{vE}\over{\lambda}}\right)
  \int J\tilde\Sigma\,{\rm d}\phi
  =-{{3{\rm i}}\over{2}}\int J\tilde{\sI}\Omega^2
  {\rm e}^{{\rm i}\phi}\,{\rm d}\phi}&\nonumber\\
  &&-{\rm i}\lambda\partial_\lambda\int JR_\lambda{\sT}^{\lambda\lambda}
  \,{\rm e}^{{\rm i}\phi}\,{\rm d}\phi+
  2\partial_\lambda\int JR^2{\sT}^{\lambda\phi}
  \,{\rm e}^{{\rm i}\phi}\,{\rm d}\phi\nonumber\\
  &&-{{1}\over{\lambda}}\int JR^2{\sT}^{\lambda\phi}
  ({\rm e}^{{\rm i}\phi}+E-\lambda E')\,{\rm d}\phi\nonumber\\
  &&-{\rm i}\int JR^2{\sT}^{\phi\phi}\,{\rm e}^{{\rm i}\phi}\,{\rm d}\phi
  \nonumber\\
  &&+\int J\tilde\Sigma(2R^2f^\phi_0-{\rm i}\lambda R_\lambda f^\lambda_0)
  \,{\rm e}^{{\rm i}\phi}\,{\rm d}\phi.
\end{eqnarray}
Thus a complete set of one-dimensional equations has been obtained
that determine the evolution of the mass, angular momentum and
eccentricity vector due to internal stresses and external forcing.  It
remains to evaluate the equations explicitly for the case of a
Maxwellian viscoelastic model with an alpha viscosity.

\subsection{Vertical structure and vertical motion}

The shear tensor of the orbital motion will be required below.  This
is defined in general by
\begin{equation}
  S^{ab}={\textstyle{{1}\over{2}}}(\nabla^au^b+\nabla^bu^a),
\end{equation}
and has the expansion
\begin{equation}
  S^{ab}=S^{ab}_0+O(\epsilon).
\end{equation}
The horizontal components at leading order are
\begin{equation}
  S^{\lambda\lambda}_0={{(R^3R_{\lambda\phi}+RR_{\lambda\phi}R_\phi^2+
  R_\lambda R_\phi^3-RR_\lambda R_\phi R_{\phi\phi})\Omega}\over
  {R^3R_\lambda^3}},
\end{equation}
\begin{equation}
  S^{\lambda\phi}_0={{(R^2+R_\phi^2)\partial_\lambda\Omega+
  (R_\lambda R_{\phi\phi}-R_\phi R_{\lambda\phi})\Omega}\over
  {2R^2R_\lambda^2}},
\end{equation}
\begin{equation}
  S^{\phi\phi}_0=-{{R_\phi}\over{R^3R_\lambda}}(R\partial_\lambda\Omega+
  R_\lambda\Omega),
\end{equation}
where equation (\ref{dphir2om}) has been used.  The orbital
contribution to the divergence $\nabla_au^a$ is
\begin{equation}
  \Delta={{1}\over{J}}\partial_\phi(J\Omega)=
  \left({{GM}\over{\lambda^3}}\right)^{1/2}g_1,
\end{equation}
where $g_1(\theta;\lambda,T)$ is a dimensionless quantity, equal to
zero for a circular disc (see Appendix~A).
Note, however, that there is also a vertical shear component
$S^{zz}_0=\partial_\zeta u^z_1$ of the same order, which remains to be
evaluated.  This contributes to both the divergence and the
dissipation rate at leading order.

Noting that
\begin{equation}
  \int\,{{{\rm d}\phi}\over{\Omega}}=P=
  2\pi(1-e^2)^{-3/2}\left({{\lambda^3}\over{GM}}\right)^{1/2}
\end{equation}
is the orbital period, one may write
\begin{equation}
  \tilde\Sigma=\Sigma\,g_2,
\end{equation}
where $g_2(\theta;\lambda,T)$ is a second dimensionless quantity,
equal to unity for a circular disc, and satisfying
\begin{equation}
  (1+e\cos\theta)^2\partial_\theta\ln g_2=-g_1.
\end{equation}
Further useful quantities are defined according to
\begin{equation}
  \lambda\partial_\lambda\Omega=
  \left({{GM}\over{\lambda^3}}\right)^{1/2}\,g_3,
\end{equation}
\begin{equation}
  \partial_\phi\Omega=\left({{GM}\over{\lambda^3}}\right)^{1/2}\,g_4
\end{equation}
and
\begin{equation}
  S^{ab}_0=\left({{GM}\over{\lambda^3}}\right)^{1/2}\,s^{ab},
\end{equation}
so that $(g_3,g_4,s^{\lambda\lambda},\lambda
s^{\lambda\phi},\lambda^2s^{\phi\phi})$ are dimensionless.  These
quantities are written out in Appendix~A.

The equation of mass conservation (\ref{drho}) at leading order
[$O(\epsilon^s)$] is
\begin{equation}
  (\Omega\partial_\phi+u^z_1\partial_\zeta)\rho_0=
  -\rho_0(\Delta+\partial_\zeta u^z_1).
  \label{rho0b}
\end{equation}
The energy equation (\ref{dp}) at leading order [$O(\epsilon^{s+2})$] is
\begin{eqnarray}
  \lefteqn{\left({{1}\over{\gamma-1}}\right)
  (\Omega\partial_\phi+u^z_1\partial_\zeta)p_0=
  -\left({{\gamma}\over{\gamma-1}}\right)p_0
  (\Delta+\partial_\zeta u^z_1)}&\nonumber\\
  &&+T^{\lambda\lambda}_0S_{\lambda\lambda0}+
  2T^{\lambda\phi}_0S_{\lambda\phi0}+T^{\phi\phi}_0S_{\phi\phi0}+
  T^{zz}_0\partial_\zeta u^z_1\nonumber\\
  &&-\partial_\zeta F_0.
  \label{p0}
\end{eqnarray}
The vertical component of the equation of motion (\ref{du}) at
leading order [$O(\epsilon^{s+1})$] is
\begin{equation}
  \rho_0(\Omega\partial_\phi+u^z_1\partial_\zeta)u^z_1=
  -\rho_0\Phi_2\zeta-\partial_\zeta p_0+
  \partial_\zeta T^{zz}_0.
  \label{w1}
\end{equation}
The required components of the stress equation (\ref{tab}) at leading
order [$O(\epsilon^{s+2})$] are
\begin{eqnarray}
  \lefteqn{T^{\lambda\lambda}_0+\tau\left[
  (\Omega\partial_\phi+u^z_1\partial_\zeta)T^{\lambda\lambda}_0+
  2(\Delta+\partial_\zeta u^z_1)T^{\lambda\lambda}_0\right]}&\nonumber\\
  &&=2\mu_0S^{\lambda\lambda}_0+
  (\mu_{{\rm b}0}-{\textstyle{{2}\over{3}}}\mu_0)
  (\Delta+\partial_\zeta u^z_1)g^{\lambda\lambda},
  \label{tll}
\end{eqnarray}
\begin{eqnarray}
  \lefteqn{T^{\lambda\phi}_0+\tau\left[
  (\Omega\partial_\phi+u^z_1\partial_\zeta)T^{\lambda\phi}_0-
  T^{\lambda\lambda}_0\partial_\lambda\Omega-
  T^{\lambda\phi}_0\partial_\phi\Omega\right.}&\nonumber\\
  &&\left.+2(\Delta+\partial_\zeta u^z_1)T^{\lambda\phi}_0\right]\nonumber\\
  &&=2\mu_0S^{\lambda\phi}_0+
  (\mu_{{\rm b}0}-{\textstyle{{2}\over{3}}}\mu_0)
  (\Delta+\partial_\zeta u^z_1)g^{\lambda\phi},
\end{eqnarray}
\begin{eqnarray}
  \lefteqn{T^{\phi\phi}_0+\tau\left[
  (\Omega\partial_\phi+u^z_1\partial_\zeta)T^{\phi\phi}_0-
  2T^{\lambda\phi}_0\partial_\lambda\Omega-
  2T^{\phi\phi}_0\partial_\phi\Omega\right.}&\nonumber\\
  &&\left.+2(\Delta+\partial_\zeta u^z_1)T^{\phi\phi}_0\right]\nonumber\\
  &&=2\mu_0S^{\phi\phi}_0+
  (\mu_{{\rm b}0}-{\textstyle{{2}\over{3}}}\mu_0)
  (\Delta+\partial_\zeta u^z_1)g^{\phi\phi},
\end{eqnarray}
\begin{eqnarray}
  \lefteqn{T^{zz}_0+\tau\left[
  (\Omega\partial_\phi+u^z_1\partial_\zeta)T^{zz}_0+
  2\Delta T^{zz}_0\right]}&\nonumber\\
  &&=2\mu_0\partial_\zeta u^z_1+
  (\mu_{{\rm b}0}-{\textstyle{{2}\over{3}}}\mu_0)
  (\Delta+\partial_\zeta u^z_1).
  \label{tzz}
\end{eqnarray}
The constitutive relations at leading
order are
\begin{equation}
  F_0=-{{16\sigma T_0^{3-y}}\over{3C_\kappa\rho_0^{1+x}}}\partial_\zeta T_0,
  \label{f0}
\end{equation}
\begin{equation}
  p_0={{k\rho_0T_0}\over{\mu_{\rm m}m_{\rm H}}},
\end{equation}
\begin{equation}
  \mu_0=\alpha p_0\left({{GM}\over{\lambda^3}}\right)^{-1/2},\qquad
  \mu_{{\rm b}0}=\alpha_{\rm b}p_0\left({{GM}\over{\lambda^3}}\right)^{-1/2}.
  \label{mu0}
\end{equation}

The solution of equations (\ref{rho0b})--(\ref{mu0}) is found by a {\it
  non-linear separation of variables}, a similar method to that used
for warped discs (Ogilvie 2000).  As an intermediate step, one
proposes that the solution should satisfy the {\it generalized
  vertical equilibrium\/} relations
\begin{eqnarray}
  {{\partial p_0}\over{\partial\zeta}}&=&-f_2\,
  \rho_0\left({{GM}\over{\lambda^3}}\right)\zeta,
  \label{dpdz}\\
  {{\partial F_0}\over{\partial\zeta}}&=&f_1\,
  {{9\alpha}\over{4}}p_0\left({{GM}\over{\lambda^3}}\right)^{1/2},
\end{eqnarray}
in addition to equations (\ref{f0})--(\ref{mu0}).  Here
$f_1(\theta;\lambda,T)$ and $f_2(\theta;\lambda,T)$ are dimensionless
functions to be determined, and which are equal to unity for a
circular disc.  Physically, $f_1$ differs from unity in an eccentric
disc because of the enhanced dissipation of energy and because of
compressive heating and cooling (equation \ref{p0}).  Similarly, $f_2$
reflects changes in the usual hydrostatic vertical equilibrium
resulting from the vertical velocity, the vertical viscous stress and
the variation of the vertical oscillation frequency around the orbit
(equation \ref{w1}).  Following Ogilvie (2000), one solves these
equations by reducing them to a standard dimensionless form.  One
identifies a natural physical unit for the thickness of the disc,
\begin{eqnarray}
  \lefteqn{U_H=\left({{9\alpha}\over{4}}\right)^{1/(6+x-2y)}
  \Sigma^{(2+x)/(6+x-2y)}
  \left({{GM}\over{\lambda^3}}\right)^{-(5-2y)/2(6+x-2y)}}&\nonumber\\
  &&\times\left({{\mu_{\rm m}m_{\rm H}}\over{k}}\right)^{-(4-y)/(6+x-2y)}
  \left({{16\sigma}\over{3C_\kappa}}\right)^{-1/(6+x-2y)},
\end{eqnarray}
and for the other variables according to
\begin{equation}
  U_\rho=\Sigma U_H^{-1},
\end{equation}
\begin{equation}
  U_p=\left({{GM}\over{\lambda^3}}\right)U_H^2U_\rho,
\end{equation}
\begin{equation}
  U_T=\left({{GM}\over{\lambda^3}}\right)
  \left({{\mu_{\rm m}m_{\rm H}}\over{k}}\right)U_H^2,
\end{equation}
\begin{equation}
  U_F=\left({{9\alpha}\over{4}}\right)
  \left({{GM}\over{\lambda^3}}\right)^{1/2}U_HU_p.
\end{equation}
The solution of the generalized vertical equilibrium equations is then
\begin{eqnarray}
  \lefteqn{\zeta=\zeta_*\,f_1^{1/(6+x-2y)}f_2^{-(3-y)/(6+x-2y)}}&\nonumber\\
  &&\times g_2^{(2+x)/(6+x-2y)}U_H,
\end{eqnarray}
\begin{eqnarray}
  \lefteqn{\rho_0=\rho_*(\zeta_*)\,f_1^{-1/(6+x-2y)}f_2^{(3-y)/(6+x-2y)}}
  &\nonumber\\
  &&\times g_2^{2(2-y)/(6+x-2y)}U_\rho,
\end{eqnarray}
\begin{eqnarray}
  \lefteqn{p_0=p_*(\zeta_*)\,f_1^{1/(6+x-2y)}f_2^{(3+x-y)/(6+x-2y)}}
  &\nonumber\\
  &&\times g_2^{2(4+x-y)/(6+x-2y)}U_p,
\end{eqnarray}
\begin{eqnarray}
  \lefteqn{T_0=T_*(\zeta_*)\,f_1^{2/(6+x-2y)}f_2^{x/(6+x-2y)}}&\nonumber\\
  &&\times g_2^{2(2+x)/(6+x-2y)}U_T,
\end{eqnarray}
\begin{eqnarray}
  \lefteqn{F_0=F_*(\zeta_*)\,f_1^{(8+x-2y)/(6+x-2y)}f_2^{x/(6+x-2y)}}
  &\nonumber\\
  &&\times g_2^{(10+3x-2y)/(6+x-2y)}U_F,
\end{eqnarray}
where the starred variables satisfy the dimensionless ODEs
\begin{eqnarray}
  {{{\rm d}p_*}\over{{\rm d}\zeta_*}}&=&-\rho_*\zeta_*,\\
  {{{\rm d}F_*}\over{{\rm d}\zeta_*}}&=&p_*,\\
  {{{\rm d}T_*}\over{{\rm d}\zeta_*}}&=&-\rho_*^{1+x}T_*^{-3+y}F_*,\\
  p_*&=&\rho_*T_*,
\end{eqnarray}
subject to the boundary conditions
\begin{equation}
  F_*(0)=\rho_*(\zeta_{{\rm s}*})=T_*(\zeta_{{\rm s}*})=0
\end{equation}
and the normalization of surface density,
\begin{equation}
  \int_{-\zeta_{{\rm s}*}}^{\zeta_{{\rm s}*}}\rho_*\,{\rm d}\zeta_*=1.
\end{equation}
Here $\zeta_{{\rm s}*}$ is the dimensionless height of the upper
surface of the disc.  Also required is the dimensionless second
vertical moment of the density,
\begin{equation}
  {\sI}_*=\int_{-\zeta_{{\rm s}*}}^{\zeta_{{\rm s}*}}\rho_*\zeta_*^2\,
  {\rm d}\zeta_*.
\end{equation}
The solution of these dimensionless ODEs depends only on the opacity
law and is easily obtained numerically (Ogilvie 2000).\footnote{It is
  assumed here that the disc is highly optically thick so that `zero
  boundary conditions' are adequate.  The corrections for finite
  optical depth are described by Ogilvie (2000).} For Thomson opacity
one finds $\zeta_{{\rm s}*}\approx2.383$ and ${\sI}_*=0.6777$.  For
Kramers opacity one finds $\zeta_{{\rm s}*}\approx2.543$ and
${\sI}_*=0.7094$.

One further proposes that the vertical velocity is of the form
\begin{equation}
  u^z_1=f_3\,\left({{GM}\over{\lambda^3}}\right)^{1/2}\zeta,
\end{equation}
where $f_3(\theta;\lambda,T)$ is a third dimensionless function, equal
to zero for a circular disc, and that the stress components at leading
order are of the form
\begin{equation}
  T^{ab}_0=t^{ab}(\theta;\lambda,T)\,p_0,
\end{equation}
so that $(t^{\lambda\lambda},\lambda
t^{\lambda\phi},\lambda^2t^{\phi\phi},t^{zz})$ are dimensionless
coefficients.

When these tentative solutions are substituted into equations
(\ref{rho0b})--(\ref{tzz}) one obtains a number of dimensionless ODEs in
$\theta$, which must be satisfied if the solution is to be valid.  From
equation (\ref{rho0b}) one obtains
\begin{eqnarray}
  \lefteqn{(1+e\cos\theta)^2\left[-\partial_\theta\ln f_1+
  (3-y)\partial_\theta\ln f_2\right]}&\nonumber\\
  &&=-(6+x-2y)f_3-(2+x)g_1.
\end{eqnarray}
Similarly, from equation (\ref{p0}) one obtains
\begin{eqnarray}
  \lefteqn{(1+e\cos\theta)^2\left({{1}\over{\gamma-1}}\right)
  \partial_\theta\ln f_2=-\left({{\gamma+1}\over{\gamma-1}}\right)f_3
  -g_1}&
  \nonumber\\
  &&+t^{\lambda\lambda}s_{\lambda\lambda}+
  2t^{\lambda\phi}s_{\lambda\phi}+t^{\phi\phi}s_{\phi\phi}+t^{zz}f_3-
  {{9}\over{4}}\alpha f_1.
\end{eqnarray}
From equation (\ref{w1}) one obtains
\begin{equation}
  (1+e\cos\theta)^2\partial_\theta f_3=-f_3^2-(1+e\cos\theta)^3+
  f_2(1-t^{zz}).
\end{equation}
Finally, from equations (\ref{tll})--(\ref{tzz}) one obtains
\begin{eqnarray}
  \lefteqn{t^{\lambda\lambda}+{\rm We}\left[
  (1+e\cos\theta)^2(\partial_\theta t^{\lambda\lambda}+
  t^{\lambda\lambda}\partial_\theta\ln f_2)+
  (3f_3+g_1)t^{\lambda\lambda}\right]}&\nonumber\\
  &&=2\alpha s^{\lambda\lambda}+
  (\alpha_{\rm b}-{\textstyle{{2}\over{3}}}\alpha)(g_1+f_3)g^{\lambda\lambda},
\end{eqnarray}
\begin{eqnarray}
  \lefteqn{\lambda t^{\lambda\phi}+{\rm We}\left[
  (1+e\cos\theta)^2(\partial_\theta(\lambda t^{\lambda\phi})+
  \lambda t^{\lambda\phi}\partial_\theta\ln f_2)\right.}&\nonumber\\
  &&\left.+(3f_3+g_1)\lambda t^{\lambda\phi}-g_3t^{\lambda\lambda}-
  g_4\lambda t^{\lambda\phi}\right]\nonumber\\
  &&=2\alpha\lambda s^{\lambda\phi}+
  (\alpha_{\rm b}-{\textstyle{{2}\over{3}}}\alpha)(g_1+f_3)
  \lambda g^{\lambda\phi},
\end{eqnarray}
\begin{eqnarray}
  \lefteqn{\lambda^2t^{\phi\phi}+{\rm We}\left[
  (1+e\cos\theta)^2(\partial_\theta(\lambda^2t^{\phi\phi})+
  \lambda^2t^{\phi\phi}\partial_\theta\ln f_2)\right.}&\nonumber\\
  &&\left.+(3f_3+g_1)\lambda^2t^{\phi\phi}-2g_3\lambda t^{\lambda\phi}-
  2g_4\lambda^2t^{\phi\phi}\right]\nonumber\\
  &&=2\alpha\lambda^2s^{\phi\phi}+
  (\alpha_{\rm b}-{\textstyle{{2}\over{3}}}\alpha)(g_1+f_3)
  \lambda^2g^{\phi\phi},
\end{eqnarray}
\begin{eqnarray}
  \lefteqn{t^{zz}+{\rm We}\left[
  (1+e\cos\theta)^2(\partial_\theta t^{zz}+
  t^{zz}\partial_\theta\ln f_2)+(f_3+g_1)t^{zz}\right]}&\nonumber\\
  &&=2\alpha f_3+
  (\alpha_{\rm b}-{\textstyle{{2}\over{3}}}\alpha)(g_1+f_3).
\end{eqnarray}

These ODEs should be solved for the functions
$(f_1,f_2,f_3,t^{\lambda\lambda},\lambda
t^{\lambda\phi},\lambda^2t^{\phi\phi},t^{zz})$ subject to periodic
boundary conditions $f_1(2\pi;\lambda,T)=f_1(0;\lambda,T)$, etc.  Note
that, in the case ${\rm We}=0$, the equations for $t^{ab}$ become
algebraic, reducing the problem to third order.  Note also that, in
the limit of a circular disc ($e=0$, $g_1=0$, $g_2=1$,
$g_3=-{\textstyle{{3}\over{2}}}$, $g_4=0$, $g^{\lambda\lambda}=1$,
$\lambda g^{\lambda\phi}=0$, $\lambda^2g^{\phi\phi}=1$,
$s_{\lambda\lambda}=0$,
$\lambda^{-1}s_{\lambda\phi}=-{\textstyle{{3}\over{4}}}$,
$\lambda^{-2}s_{\phi\phi}=0$), the solution is $f_1=1$, $f_2=1$,
$f_3=0$, $t^{\lambda\lambda}=0$, $\lambda
t^{\lambda\phi}=-{\textstyle{{3}\over{2}}}\alpha$,
$\lambda^2t^{\phi\phi}={\textstyle{{9}\over{2}}}{\rm We}\,\alpha$,
$t^{zz}=0$.

\subsection{Evaluation of the stress integrals}

The evolutionary equations may be written in the form
\begin{equation}
  \dot\Sigma+{{1}\over{\lambda}}
  \partial_\lambda(\lambda v\Sigma)=0,
  \label{fundamental_mass}
\end{equation}
\begin{equation}
  \Sigma v{{{\rm d}h}\over{{\rm d}\lambda}}=
  {{1}\over{\lambda}}\partial_\lambda
  \left(Q_1{\sI}{{GM}\over{\lambda}}\right)+
  {{\Sigma}\over{P}}\int R^2f^\phi_0\,{{{\rm d}\phi}\over{\Omega}},
  \label{fundamental_am}
\end{equation}
\begin{eqnarray}
  \lefteqn{\Sigma h\left(\dot E+vE'-{{vE}\over{\lambda}}\right)=
  \partial_\lambda\left(Z_1{\sI}{{GM}\over{\lambda^2}}\right)
  +Z_2{\sI}{{GM}\over{\lambda^3}}}&\nonumber\\
  &&+{{\Sigma}\over{P}}\int
  (2R^2f^\phi_0-{\rm i}\lambda R_\lambda f^\lambda_0)
  \,{\rm e}^{{\rm i}\phi}\,{{{\rm d}\phi}\over{\Omega}},
  \label{fundamental_e}
\end{eqnarray}
where $Q_1$ is a dimensionless real coefficient defined by
\begin{equation}
  \int JR^2{\sT}^{\lambda\phi}\,{\rm d}\phi=
  Q_1{{GM}\over{\lambda^2}}\int J\tilde{\sI}\,{\rm d}\phi,
\end{equation}
and $Z_1$ and $Z_2$ are dimensionless complex coefficients defined by
\begin{equation}
  {{1}\over{\lambda}}\int J
  (2R^2{\sT}^{\lambda\phi}-{\rm i}\lambda R_\lambda{\sT}^{\lambda\lambda})
  \,{\rm e}^{{\rm i}\phi}\,{\rm d}\phi=
  Z_1{{GM}\over{\lambda^3}}\int J\tilde{\sI}\,{\rm d}\phi,
\end{equation}
\begin{eqnarray}
  \lefteqn{-{{3{\rm i}}\over{2}}\int J\tilde{\sI}\Omega^2
  {\rm e}^{{\rm i}\phi}\,{\rm d}\phi+
  {{1}\over{\lambda}}\int JR^2{\sT}^{\lambda\phi}
  ({\rm e}^{{\rm i}\phi}-E+\lambda E')\,{\rm d}\phi}&\nonumber\\
  &&-{\rm i}\int JR^2{\sT}^{\phi\phi}{\rm e}^{{\rm i}\phi}
  \,{\rm d}\phi=
  Z_2{{GM}\over{\lambda^3}}\int J\tilde{\sI}\,{\rm d}\phi.
\end{eqnarray}

To evaluate the stress integrals, first note the relation
\begin{equation}
  \int p_0\,{\rm d}\zeta=f_2\left({{GM}\over{\lambda^3}}\right)\tilde{\sI},
\end{equation}
which follows from equation (\ref{dpdz}) after an integration by
parts.  Then
\begin{equation}
  \sT^{ab}=(t^{ab}-g^{ab})f_2\left({{GM}\over{\lambda^3}}\right)\tilde{\sI}.
\end{equation}
Define the dimensionless function $f_4(\theta;\lambda,T)$ by
\begin{equation}
  J\tilde{\sI}=f_4\lambda{\sI},
\end{equation}
so that
\begin{equation}
  f_4={{f_5}\over{{{1}\over{2\pi}}\int f_5\,{\rm d}\theta}},
\end{equation}
where
\begin{eqnarray}
  \lefteqn{f_5=(1+e\cos\theta)^{-2}f_1^{2/(6+x-2y)}f_2^{-2(3-y)/(6+x-2y)}}
  &\nonumber\\
  &&\qquad\times g_2^{2(2+x)/(6+x-2y)}.
\end{eqnarray}
Then one finds
\begin{equation}
  Q_1={{1}\over{2\pi}}\int f_2f_4(1+e\cos\theta)^{-2}
  \lambda(t^{\lambda\phi}-g^{\lambda\phi})\,{\rm d}\theta,
\end{equation}
\begin{eqnarray}
  \lefteqn{Z_1={{1}\over{2\pi}}{\rm e}^{{\rm i}\omega}
  \int f_2f_4\left[2(1+e\cos\theta)^{-2}\lambda
  (t^{\lambda\phi}-g^{\lambda\phi})\right.}&\nonumber\\
  &&\left.\qquad\qquad
  -{\rm i}R_\lambda(t^{\lambda\lambda}-g^{\lambda\lambda})\right]
  {\rm e}^{{\rm i}\theta}\,{\rm d}\theta,
\end{eqnarray}
\begin{eqnarray}
  \lefteqn{Z_2={{1}\over{2\pi}}(\lambda E'-E)
  \int f_2f_4(1+e\cos\theta)^{-2}
  \lambda(t^{\lambda\phi}-g^{\lambda\phi})\,{\rm d}\theta}&\nonumber\\
  &&+{{1}\over{2\pi}}{\rm e}^{{\rm i}\omega}
  \int\left\{-{{3{\rm i}}\over{2}}f_4(1+e\cos\theta)^4+
  f_2f_4(1+e\cos\theta)^{-2}\right.\nonumber\\
  &&\left.\qquad
  \times\left[\lambda(t^{\lambda\phi}-g^{\lambda\phi})-
  {\rm i}\lambda^2(t^{\phi\phi}-g^{\phi\phi})\right]
  \right\}{\rm e}^{{\rm i}\theta}\,{\rm d}\theta.
\end{eqnarray}

Finally, a relation is required between $\Sigma$ and ${\sI}$.  One finds
\begin{eqnarray}
  \lefteqn{{\sI}=Q_2C_{\sI}\alpha^{2/(6+x-2y)}
  \left({{GM}\over{\lambda^3}}\right)^{-(5-2y)/(6+x-2y)}}&\nonumber\\
  &&\times\Sigma^{(10+3x-2y)/(6+x-2y)},
\end{eqnarray}
where
\begin{eqnarray}
  \lefteqn{C_{\sI}=\left({{9}\over{4}}\right)^{2/(6+x-2y)}{\sI}_*
  \left({{\mu_{\rm m}m_{\rm H}}\over{k}}\right)^{-2(4-y)/(6+x-2y)}}&\nonumber\\
  &&\times\left({{16\sigma}\over{3C_\kappa}}\right)^{-2/(6+x-2y)}
\end{eqnarray}
is a constant and $Q_2$ is a dimensionless coefficient defined by
\begin{equation}
  Q_2={{1}\over{2\pi}}(1-e^2)^{3/2}\int f_5\,{\rm d}\theta,
\end{equation}
and equal to unity for a circular disc.  The dimensionless
coefficients can all be evaluated numerically from the solution of the
dimensionless ODEs in Section~4.4.  For given values of the model
parameters $(\alpha,\alpha_{\rm b},{\rm We},\gamma,x,y)$, the
quantities $(Q_1,Q_2,Z_1/{\rm e}^{{\rm i}\omega},Z_2/{\rm e}^{{\rm
    i}\omega})$ depend only on the dimensionless dynamical variables
$(e,\lambda e',\lambda e\omega')$.

\subsection{Relation to the Gauss perturbation equations}

It may now be shown how the evolutionary equations for a continuous
disc relate to those derived in Section~1.2 for a test body.  The
relation between the ordinary cylindrical polar components $(f_{\hat
  R},f_{\hat\phi})$ of the perturbing force and the contravariant
orbital components $(f^\lambda,f^\phi)$ is
\begin{equation}
  f_{\hat R}=R_\lambda f^\lambda+R_\phi f^\phi,
\end{equation}
\begin{equation}
  f_{\hat\phi}=Rf^\phi.
\end{equation}
Thus equation (\ref{gauss_h}) for the test body becomes
\begin{equation}
  {{{\rm d}h}\over{{\rm d}t}}=v{{{\rm d}h}\over{{\rm d}\lambda}}=R^2f^\phi,
\end{equation}
where $v={\rm d}\lambda/{\rm d}t$.  This is equivalent to equation
(\ref{compare1}) for the disc, except that the disc equation involves
a time-average over the orbit, and the perturbing force includes
contributions from internal stresses.

Similarly, equation (\ref{gauss_e}) for the test body may be (after
some algebra) brought into the form
\begin{equation}
  h\left({{{\rm d}E}\over{{\rm d}t}}-{{vE}\over{\lambda}}\right)=
  (2R^2f^\phi-{\rm i}\lambda R_\lambda f^\lambda)\,{\rm e}^{{\rm i}\phi},
\end{equation}
which is consistent, in the same sense, with equation
(\ref{compare2}).

\subsection{Relation to standard accretion disc theory}

In the absence of an external torque, equations
(\ref{fundamental_mass}) and (\ref{fundamental_am}) may be combined
into a single evolutionary equation for the surface density,
\begin{equation}
  \dot\Sigma={{3}\over{\lambda}}\partial_\lambda
  \left\{\lambda^{1/2}\partial_\lambda
  \left[\lambda^{1/2}\left(-{{2Q_1}\over{3}}\right){\sI}
  \left({{GM}\over{\lambda^3}}\right)^{1/2}\right]\right\}.
\end{equation}
In the limit of a circular disc, $\lambda$ reduces to $R$ and one
finds that $Q_1\to-3\alpha/2$.  One then obtains the standard
diffusion equation for the surface density of a circular accretion
disc (e.g. Pringle 1981),
\begin{equation}
  \dot\Sigma={{3}\over{R}}\partial_R
  \left[R^{1/2}\partial_R\left(R^{1/2}\nu\Sigma\right)\right],
\end{equation}
where $\nu=\alpha\Omega({\sI}/\Sigma)$ is the vertically averaged
kinematic viscosity.  In an eccentric disc, therefore, the surface
density diffuses viscously in much the same way as in a circular disc,
except that it is best considered as diffusing in the space of
semi-latus rectum $\lambda$ (which is equivalent to specific angular
momentum).  Non-linear couplings arise, however, because the
coefficient $Q_1$ depends on the local eccentricity distribution,
specifically on $e$, $\lambda e'$ and $\lambda e\omega'$.  The
elliptical distortion of the orbits modifies the stresses that leads
to accretion.  The eccentricity obeys its own equation
(\ref{fundamental_e}), which has the character of a complex,
non-linear advection-diffusion equation.

\section{Linear theory}

In general the dimensionless ODEs must be solved numerically and the
$Q$- and $Z$-coefficients determined from the solution.  However, for
small eccentricities the equations can be solved by expanding in
powers of $e$.  In doing so it is assumed that $\lambda E'$ is $O(e)$,
i.e. that the length-scale on which $e$ or $\omega$ varies is
comparable to the radius of the disc.

One then finds
\begin{eqnarray}
  Q_1&=&-{{3}\over{2}}\alpha+O(e^2),\\
  Q_2&=&1+O(e^2),\\
  Z_1&=&c_1E+c_2\lambda E'+O(e^3),\\
  Z_2&=&c_3E+c_4\lambda E'+O(e^3).
\end{eqnarray}
The complex coefficients $c_i$ depend only on the dimensionless
parameters $(\alpha,\alpha_{\rm b},{\rm We},\gamma,x,y)$, and can be
obtained by a straightforward algebraic calculation.  Such a
calculation is not presented here because the resulting expressions
are too complicated to write down, and it is no more difficult to
evaluate the dimensionless coefficients numerically in the non-linear
case.  The only simple, general statement that can be made is that the
$c_i$ become purely imaginary in the inviscid limit
$\alpha,\alpha_{\rm b}\to0$.

If the higher-order terms are neglected, and in the absence of
external forces, one obtains a complex linear diffusion-type equation
for the eccentricity, together with the standard evolutionary equation
for the surface density, which is decoupled from the eccentricity.
This is similar to the situation in Section~2 except that
three-dimensional and radiative effects have been included.  Of
greatest interest is the coefficient $c_2$, because the small-scale
instability discussed in Section~2 exists if ${\rm Re}(c_2)<0$.

For the purposes of illustration, suppose that $\alpha=0.1$ and
$\gamma=5/3$.  In a purely viscous disc with ${\rm We}=0$, it is then
found that stability requires a bulk viscosity $\alpha_{\rm b}>0.350$
in the case of Thomson opacity, or $\alpha_{\rm b}>0.176$ in the case
of Kramers opacity.  It is by no means clear that such values are
realistic.  However, the instability is easily quenched when a
non-zero relaxation time is introduced.  Even when $\alpha_{\rm b}=0$,
the instability is absent for $0.398<{\rm We}<2.294$ (Thomson) or
$0.333<{\rm We}<2.467$ (Kramers).  This contrasts with the oversimple
model of Section~2, which generally predicted narrower stability
intervals and suggested that the instability persisted for all ${\rm
  We}$ when $\mu_{\rm b}=0$.  These results emphasize the importance
of including three-dimensional effects and radiative damping when
treating eccentric discs.

One of the features of the linear analysis is that shows that vertical
motion is always present in an eccentric disc.  To understand this,
consider a simplified example in which the gas is isothermal so that
the pressure is proportional to the density.  Suppose further that the
eccentricity is instantaneously uniform.  In the absence of vertical
motion the velocity divergence is zero and the density and pressure
must then be constant along any orbit, for any value of $z$.  However,
this is incompatible with hydrostatic vertical equilibrium because the
vertical gravitational acceleration varies around any eccentric orbit.
Of course, this effect is missing in two-dimensional treatments of
eccentric discs.

\section{An illustrative non-linear calculation}

The practical nature of the non-linear evolutionary equations is now
demonstrated by a simple illustrative calculation.  In any detailed
application, careful consideration must be given to the formulation of
appropriate boundary conditions and to any sources of mass and/or
eccentricity to be added to the equations.  At a free boundary of the
disc, where $\Sigma\to0$, equation (\ref{fundamental_e}) generally has
a singular point and one must select the regular solution.  If the
disc is terminated by an external agency, however, a different
boundary condition on the eccentricity may apply.  These issues should
be addressed in future work, within the context of specific
applications.  For the present purposes, it is convenient to study a
simple test problem with idealized boundary conditions.

The parameters adopted are $\alpha=0.1$, $\alpha_{\rm b}=0$, ${\rm
  We}=0.5$, $\gamma=5/3$, $x=1$ and $y=-7/2$, appropriate to a disc in
which Kramers opacity is dominant.  (The corresponding linear-theory
coefficients are $c_1=0.075+0.972{\rm i}$, $c_2=0.030+0.631{\rm i}$,
$c_3=-0.228+0.955{\rm i}$ and $c_4=-0.198-0.222{\rm i}$.  Therefore
the short-wavelength instability is absent.)  For this opacity law
there exists a self-similar solution with
$\Sigma\propto\lambda^{-3/4}$ and $E=0$, representing a steady,
circular accretion disc having an arbitrarily small inner radius.
According to the theories of Syer \& Clarke (1992, 1993) and Lyubarskij et
al. (1994), if such a disc is made uniformly eccentric, with all the
orbits aligned, it should remain so.  In order to contrast the results
of the present theory with these earlier calculations, the non-linear
evolutionary equations (\ref{fundamental_mass})--(\ref{fundamental_e})
are solved starting from an initial condition with surface density
$\Sigma\propto\lambda^{-3/4}$ and uniform eccentricity $E=0.3$.  The
equations are solved in a finite domain, $1<\lambda<100$, with
boundary conditions $\partial\ln\Sigma/\partial\ln\lambda=-{3/4}$ and
$\partial_\lambda E=0$ at the inner and outer edges.  These
illustrative boundary conditions are chosen to be relatively neutral
while being compatible with the notional solution proposed by Syer \&
Clarke and Lyubarskij et al.  No external forcing is applied.

The equations are first discretized in space by representing the
variables $\Sigma$ and $E$ on a set of 200 logarithmically spaced
orbits.  The derivatives are represented by simple, centred finite
differences and the scheme conserves mass exactly.  The resulting set
of temporal ODEs is solved using a fifth-order Runge-Kutta method with
adaptive step-size.  The time-step adjusts automatically to ensure
accuracy and stability of the integration.

Before the start of the run, the dimensionless coefficients
$(Q_1,Q_2,Z_1/{\rm e}^{{\rm i}\omega},Z_2/{\rm e}^{{\rm i}\omega})$
are evaluated on a rectangular grid in the space $(e,\lambda
e',\lambda e\omega')$ by solving the dimensionless ODEs of Section~4.4
in a once-for-all calculation for the chosen parameter values.  During
the run, the coefficients are then evaluated rapidly by trilinear
interpolation on the grid.

Fig.~2 shows the evolution of $e$ and $\omega$ over a time interval
$100t_\nu$, where $t_\nu=\lambda^2/\nu$ is the viscous time-scale at
the inner boundary, and $\nu=\alpha\Omega(\sI/\Sigma)$ is the
vertically averaged kinematic viscosity in a circular disc.  During
this period the surface density exhibits negligible evolution.
However, there is a rapid twisting of the disc because of differential
precession caused by the radial pressure gradient, an effect explained
in Section~2.2.  After a transient phase, the eccentricity settles
into a twisted mode that decays exponentially in time and precesses
slowly in a prograde sense.

\begin{figure}
  \centerline{\epsfbox{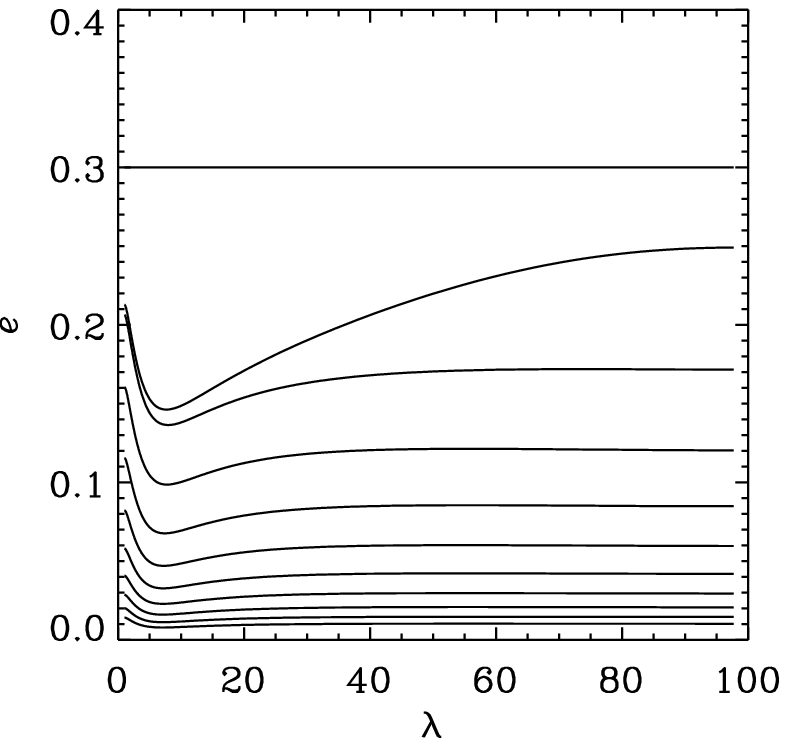}}
  \centerline{\epsfbox{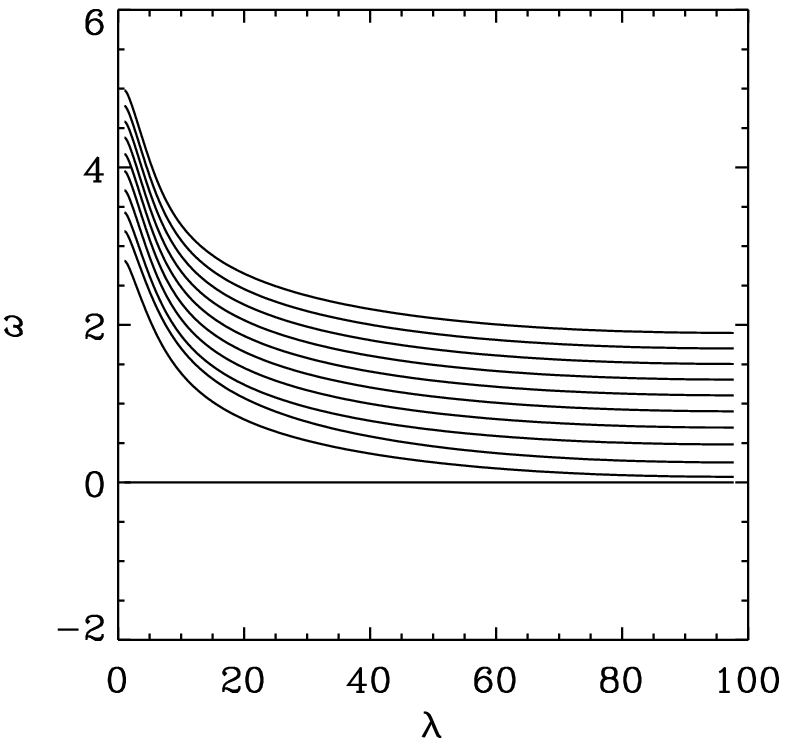}}
  \caption{Top: Evolution of the eccentricity in an initially uniformly
    eccentric disc.  The curves are ordered from top to bottom, and
    are separated by time intervals of $10t_\nu$.  Bottom: Evolution
    of the longitude of periastron (radians).  The curves are ordered
    from bottom to top.}
\end{figure}

Owing to the monotonic decay of the eccentricity, non-linear terms are
active only in the earliest stages of the evolution.  However,
non-linear effects will be critical in other applications, such as
determining the outcome of the eccentric instability in superhump
binaries.

\section{Discussion}

In this paper a comprehensive theory of eccentric accretion discs has
been presented.  Starting from the basic fluid-dynamical equations in
three dimensions, I have derived the fundamental set of
one-dimensional equations that describe how the mass, angular momentum
and eccentricity vector of a thin disc evolve as a result of internal
stresses and external forcing (equations
\ref{fundamental_mass}--\ref{fundamental_e}).  The analysis is
asymptotically exact in the limit of a thin disc, and allows for
slowly varying eccentricities of arbitrary magnitude.

These equations are generally valid and therefore of fundamental
interest.  They are the equivalent of the Gauss perturbation equations
for a continuous disc.  Previously, Lyubarskij et al. (1994) succeeded
in deriving a related set of equations for an eccentric disc by
considering the conservation of mass, angular momentum and energy.
However, their analysis is restricted to the case in which the
ellipses are all aligned and do not precess.  Their method works in
this case because a knowledge of the angular momentum and energy of an
orbiting body is sufficient to determine its semi-latus rectum and
eccentricity, but not its longitude of periastron.  A closed system of
equations is obtained only if the ellipses are artificially
constrained not to precess.  In reality, such precession is inevitable
and the evolution of the longitude of periastron must be determined
from a full analysis of the horizontal components of the equation of
motion.  This leads to an equation for the eccentricity vector, or
complex eccentricity, which is not in conservative form.

The second achievement of this paper is the explicit development of
the equations in the case of a specific stress model which, it is
hoped, gives a fair representation of the turbulent stress in an
accretion disc.  To obtain the coefficients in the evolutionary
equations requires a solution of the non-linear PDEs that govern the
azimuthal and vertical structure of the disc.  It also requires an
understanding of the relation between the turbulent stress tensor and
the velocity gradient tensor.  The simplest plausible relation,
adopted in almost all theoretical work on accretion discs, is an
effective viscosity model in which an instantaneous linear relation is
assumed, and the equation of motion therefore reduces to the
Navier-Stokes equation.  In this paper I have introduced a Maxwellian
viscoelastic model of the turbulent stress in an accretion disc.  This
generalizes the conventional alpha viscosity model to account for the
non-zero relaxation time of the turbulence, and is physically
motivated by a consideration of the nature of MHD turbulence.  The
PDEs governing the azimuthal and vertical structure of the disc,
including the effects of vertical motion, dissipation of energy and
radiative transport, have been reduced exactly to a set of
dimensionless ODEs which can be solved numerically to high accuracy to
yield the coefficients required for the evolutionary equations.  This
shows that the technique of non-linear separation of variables,
applied first to warped discs (Ogilvie 1999, 2000), is not restricted
to purely viscous models but can incorporate improved representations
of the stress as our understanding of magnetorotational turbulence
develops.

It has been confirmed that circular discs are usually viscously
unstable to short-wavelength eccentric perturbations, as found by
Lyubarskij et al. (1994), if the conventional alpha viscosity model is
adopted.  It has been noted that the instability is essentially the
same as the viscous overstability of axisymmetric modes discovered by
Kato (1978).  The instability can be suppressed by introducing a
sufficient effective bulk viscosity, although the values required may
not be realistic.  More plausibly, the instability can be suppressed
by allowing for the non-zero relaxation time of the turbulence, even
if the bulk viscosity is zero.  It has then been shown that an
initially uniformly eccentric disc does not retain its eccentricity
over many viscous time-scales, as had been suggested by Syer \& Clarke
(1992, 1993) and Lyubarskij et al. (1994).  These earlier works
neglected the differential precession caused by slightly non-Keplerian
rotation resulting from the radial pressure gradient.  This leads to
twisting of the disc, followed by viscous decay of the eccentricity.

The theory presented here goes considerably beyond previous analytical
treatments of eccentric discs.  It also provides a practical numerical
scheme that involves only one-dimensional equations and from which the
fast orbital time-scale has been eliminated.  This scheme is to be
preferred in many circumstances to a direct numerical simulation of
the fluid-dynamical equations.  Almost all direct simulations to date
attempt to represent the Navier-Stokes equation (with or without
explicit viscosity) in two dimensions.  The present analysis shows
that a two-dimensional treatment of eccentric discs may capture many
of the correct qualitative features but cannot be trusted in detail.
Vertical motion is always present in eccentric discs and radiative
damping can influence the evolution of eccentricity.  Furthermore, the
Navier-Stokes equation does not take into account the relaxation time
of the turbulence, which can be of great importance in this context.

Nevertheless, it would be valuable to make detailed comparisons
between the present theory and direct simulations (preferably
three-dimensional).  The present theory cannot be applied reliably to
thick discs, nor to situations in which the eccentricity varies
rapidly in space (i.e. on a length-scale comparable to the thickness
of the disc) or in time (i.e. on a time-scale comparable to the orbital
period).  Mean-motion resonances are therefore excluded from the
analysis, which is secular in the sense of celestial mechanics.
Nevertheless, the effect of mean-motion resonances could be included
in the evolutionary equations by adding appropriate localized source
terms for angular momentum and eccentricity.

The theory developed in this paper has much in common with the theory
of warped accretion discs (e.g. Pringle 1992; Ogilvie 1999, 2000).
One distinction, noted above, is that the equations for an eccentric
disc are not all in conservative form.  Another difference is that the
theory of warped discs is complicated by a resonance caused by the
coincidence of the orbital and epicyclic frequencies in a Keplerian
disc.  As a result, the behaviour is qualitatively different depending
on the relative magnitudes of $\alpha$ and $H/R$ (Papaloizou \& Lin
1995).  Fortunately, no such complication arises in the case of
eccentric discs.  A consistent asymptotic expansion of the
fluid-dynamical equations is possible for any value of $\alpha$, and
the fractional error in the asymptotic approximation, $O((H/R)^2)$, is
very small in most applications.

The evolutionary equations should be useful in many applications,
including understanding the eccentric planet-disc interaction and
testing theories of quasi-periodic oscillations in X-ray binaries.  In
future work the rate of change of the complex eccentricity caused by
external forcing should be evaluated explicitly in the non-linear case
for tidal forcing by a companion object on a circular or eccentric
orbit, and also for Einstein precession near a black hole.

\section*{Acknowledgments}

I acknowledge valuable discussions with Steve Lubow and with Jim
Pringle.  The idea of applying viscoelastic models to accretion discs
originated from discussions with Jim Pringle some years ago.  I thank
Scott Tremaine for helpful comments.  I acknowledge the support of
Clare College, Cambridge through a research fellowship, and the Royal
Society through a University Research Fellowship.

\appendix

\section{Geometrical quantities}

The following dimensionless quantities are required for evaluation of
the stress coefficients.  These expressions generalize those of
Lyubarskij et al. (1994) to allow for precession of the orbits, and
also correct a number of errors in that paper.  In the following, $c$
and $s$ denote $\cos\theta$ and $\sin\theta$, respectively.

Rate of change of radius with semi-latus rectum:
\begin{equation}
  R_\lambda={{1+(e-\lambda e')c-\lambda e\omega's}
  \over{(1+ec)^2}}.
\end{equation}
Inverse metric coefficients:
\begin{equation}
  g^{\lambda\lambda}={{(1+ec)^2(1+2ec+e^2)}\over
  {\left[1+(e-\lambda e')c-\lambda e\omega's\right]^2}},
\end{equation}
\begin{equation}
  \lambda g^{\lambda\phi}=-{{(1+ec)^2es}\over
  {1+(e-\lambda e')c-\lambda e\omega's}},
\end{equation}
\begin{equation}
  \lambda^2g^{\phi\phi}=(1+ec)^2.
\end{equation}
Orbital contribution to the velocity divergence:
\begin{equation}
  g_1={{(1+ec)(\lambda e's-\lambda e\omega'c-
  \lambda e^2\omega')}\over
  {1+(e-\lambda e')c-\lambda e\omega's}}.
\end{equation}
Variation of the surface density around the orbit:
\begin{equation}
  g_2={{(1-e^2)^{3/2}(1+ec)}\over
  {1+(e-\lambda e')c-\lambda e\omega's}}.
\end{equation}
Derivatives of the angular velocity:
\begin{equation}
  g_3=-{{3}\over{2}}(1+ec)^2+2(\lambda e'c+\lambda e\omega's)(1+ec),
\end{equation}
\begin{equation}
  g_4=-2es(1+ec).
\end{equation}
Orbital contribution to the shear tensor:
\begin{eqnarray}
  \lefteqn{s^{\lambda\lambda}={{(1+ec)^3}\over
  {\left[1+(e-\lambda e')c-\lambda e\omega's\right]^3}}}&\nonumber\\
  &&\times\left\{(1+ec)^2es+
  \lambda e'(1+ec+e^2s^2)s\right.\nonumber\\
  &&\left.-\lambda e\omega'\left[c+e(2+c^2)+
  e^2(4-c^2)c+e^3\right]\right\},
\end{eqnarray}
\begin{eqnarray}
  \lefteqn{\lambda s^{\lambda\phi}={{(1+ec)^3}\over
  {\left[1+(e-\lambda e')c-\lambda e\omega's\right]^2}}}&\nonumber\\
  &&\times\left\{-{{3}\over{4}}-{{7}\over{4}}ec-
   {{1}\over{4}}e^2(1+4c^2)-{{1}\over{4}}e^3c
  \right.\nonumber\\
  &&\left.+\lambda e'\left[c-{{1}\over{2}}e(1-4c^2)+
  {{1}\over{2}}e^2c\right]\right.\nonumber\\
  &&\left.+\lambda e\omega'(1+2ec+e^2)s\right\},
\end{eqnarray}
\begin{eqnarray}
  \lefteqn{\lambda^2s^{\phi\phi}={{(1+ec)^3es}\over
  {1+(e-\lambda e')c-\lambda e\omega's}}}&\nonumber\\
  &&\times\left[{{1}\over{2}}(1+ec)-
  \lambda e'c-\lambda e\omega's\right].
\end{eqnarray}
Covariant components:
\begin{eqnarray}
  \lefteqn{s_{\lambda\lambda}={{1}\over{(1+ec)^2}}
  \left\{-{{1}\over{2}}e(1+ec)s+
  \lambda e'\left(1+{{3}\over{2}}ec\right)s\right.}&\nonumber\\
  &&\left.-\lambda e\omega'
  \left[c-{{1}\over{2}}e(1-3c^2)\right]\right.\nonumber\\
  &&\left.-\lambda^2e'^2cs-\lambda^2ee'\omega'(1-2c^2)+
  \lambda^2e^2\omega'^2cs\right\},
\end{eqnarray}
\begin{eqnarray}
  \lefteqn{\lambda^{-1}s_{\lambda\phi}={{1}\over{(1+ec)^2}}}&\nonumber\\
  &&\times\left(-{{3}\over{4}}-ec+\lambda e'c+\lambda e\omega's-
  {{e^2}\over{4}}+{{1}\over{2}}\lambda ee'\right),
\end{eqnarray}
\begin{equation}
  \lambda^{-2}s_{\phi\phi}=-{{es}\over{(1+ec)^2}}.
\end{equation}

\label{lastpage}

\end{document}